\documentstyle[preprint,aps,tighten,epsf]{revtex}

\begin{document}

\newcommand{\beq}{\begin{equation}}
\newcommand{\eeq}{  \end{equation}}
\newcommand{\bea}{\begin{eqnarray}}
\newcommand{\eea}{  \end{eqnarray}}
\newcommand{\one}{\mbox{$\openone$}}

\draft

\title{Semiclassical spatial correlations in chaotic wave functions}
\author{Fabricio Toscano and Caio H. Lewenkopf}

\address{Instituto de F\'{\i}sica, 
	 Universidade do Estado do Rio de Janeiro, \\  
	 R. S\~ao Francisco Xavier 524, 20559-900 Rio de Janeiro, Brazil}

\date{\today}

\maketitle

%
\begin{abstract}
We study the spatial autocorrelation of energy eigenfunctions 
$\psi_n({\bf q})$ corresponding to classically chaotic systems 
in the semiclassical regime.
Our analysis is based on the Weyl-Wigner formalism 
for the spectral average $C_{\varepsilon}({\bf q^{+}},{\bf q^{-}},E)$ 
of $\psi_n({\bf q}^{+})\psi_n^*({\bf q}^{-})$, defined as
the average over eigenstates within an energy window $\varepsilon$
centered at $E$.
In this framework $C_{\varepsilon}$ is the Fourier transform in 
momentum space of the spectral Wigner function 
$W({\bf x},E;\varepsilon)$.
Our study reveals the chord structure that $C_{\varepsilon}$ inherits
from the spectral Wigner function 
showing the interplay between the size of the spectral average
window, 
and the spatial separation scale.
We discuss under which conditions is it possible to define a local 
system independent regime for $C_{\varepsilon}$.
In doing so, we derive an expression that bridges the existing formulae
in the literature and find expressions for $C_{\varepsilon}({\bf q^{+}},
{\bf q^{-}},E)$ valid for any separation size $|{\bf q^{+}}-{\bf q^{-}}|$.
\end{abstract}

\pacs{PACS numbers: 05.45.Mt, 03.65.Sq,73.23.-b,73.23.Ad}
%



\noindent

%
%
\section{Introduction and motivation}
\label{sec:Intro}

One of the key issues in the quantum chaos research is the quest for 
quantum fingerprints of the underlying classical dynamics of generic 
chaotic systems.
In the last decades most studies of such kind were dedicated 
to spectral properties \cite{Nato97}.
It is now established that, in general, systems with a classical chaotic 
dynamics are characterized by universal spectral fluctuations,
which is the so called Bohigas conjecture \cite{Bohigas84}. 
Much insight about this phenomenon was provided by the semiclassical 
approximation \cite{Gutzwiller90}.
Specifically, it was shown that in the semiclassical regime 
\cite{Berry85,Bogomolny96,Ozorio98} the energy level density 
autocorrelation function of a chaotic system, evaluated at energy 
separations encompassing several mean level spacings, displays 
similar statistical properties as those arising from ensembles of 
random matrices \cite{Mehta91}.
Starting from the random matrix side, advances in proving the connection 
to the spectral fluctuations of chaotic systems were also achieved 
\cite{Muzykanskii95,Andreev96}.
Although a full proof of Bohigas' conjecture is not yet available,
its domain of validity is fairly established.

Complementary to the universal view, there is another successful 
contribution of the semiclassical approach to the research in quantum 
chaos.
To every spectrum corresponding to a Hamiltonian system there are 
always deviations from the universal behavior, {\sl i.e.} system 
specific features. 
For low dimensional chaotic systems the later are usually nicely 
explained by identifying the system shortest classical periodic 
trajectories. 
For instance, knowing the actions and the stability of all
periodic orbits up to a time $T$ of a given chaotic system, the 
semiclassical approximation explains system specific spectral 
correlations within energy windows $\Delta E \ge h/T$.

The search for an universal behavior in chaotic wave functions is
more elusive.
In this case, one faces a different interplay between the quantum and
classical scales where the system specific features become important.
Actually, the investigation of non-universal signatures of the classical 
underlying dynamics in wave functions, such as scars \cite{Heller84}, 
is a most fascinating subject to which several theoretical studies were 
devoted \cite{Berry89,Kaplan98,Kaplan99,Toscano01}.
Whereas there are still several open questions to be answered,
one can naively say that it is possible to separate a ``short"
universal time from a ``long" system specific time regime in the 
description of chaotic wave functions.
This is, in a broad sense, the subject of the present paper.

The simplest statistical measure for chaotic wave functions fluctuations
is the two-point correlation function $C({\bf q^{+}},{\bf q^{-}})=
\langle \psi_n({\bf q}^{+})\psi_n^*({\bf q}^{-}) \rangle_{\bf q}$, 
where $\langle \ldots \rangle_{\bf q}$ stands for a local average 
in configuration space and $\psi_n({\bf q})$ is the wave 
function of the $n$-th energy eigenstate of a given Hamiltonian.
To the best of our knowledge this autocorrelation function was first 
discussed by Berry \cite{Berry77}, although admittedly very similar 
ideas about the characterization of chaotic wave function were already
known \cite{Shnirelman74,Voros76}.  
Berry assumed a microcanonical probability density in the classical phase 
space for chaotic quantum states and obtained a simple analytical 
expression for $C({\bf q^{+}},{\bf q^{-}})$.
In line with the statistical approach, $C$ was also reobtained 
by means of supersymmetric techniques in weakly disordered systems 
\cite{super,Gornyi01}. 
It was also numerically verified by eigenfunctions studies of 
different dynamical systems \cite{Aurich93,Li94,Backer01} and  
experimentally observed in eigenmodes of resonating microwave cavities 
\cite{Prigodin95b,Eckhardt99}.
In all such studies the agreement was always very good for spatial
separations up to a few wave lengths, {\it i.e.} ``short'' distances. 
By correcting $C$ for contributions of classical trajectories for 
large separations $|{\bf q}^+ - {\bf q}^-|$, a recent study due to 
Hortikar and Srednicki \cite{Hortikar98} improved Berry's formula for 
$C({\bf q^{+}}, {\bf q^{-}})$.

As a further motivation we add that the understanding of wave 
function correlations has some interesting direct application like, 
for instance, in the description of the  conductance fluctuations of 
quantum dots in the Coulomb blockade regime. 
\cite{Alhassid97,Vallejos99,Narimanov99,Kaplan00,Narimanov01} 
With this background in mind, we take a fresh look into the question 
of eigenstates autocorrelations using the Wigner-Weyl formalism. 
By doing it we derive a general expression for wave function spatial
correlations keeping the semiclassical approximation under strict
control. We recover as limiting cases Berry's correlation function 
\cite{Berry77} as well as the aforementioned Hortikar and Srednicki 
result \cite{Hortikar98}. 
We furthermore discuss the important role played by different kinds of 
averaging procedures and open a path for the inclusion of scar
contributions.

The structure of the paper is as follows. 
In Section \ref{sec:Wigner} we introduce the spectral Wigner function $W$, 
the object on which this study is built due to is simple relation to 
the autocorrelation function $C({\bf q^{+}},{\bf q^{-}})$.
There we also discuss aspects of the semiclassical approximation for $W$
which are essencial to understand the different limiting results 
for $C({\bf q^{+}},{\bf q^{-}})$.    
Section \ref{sec:nonuniversal} contains the main findings of this study. 
We show that the semiclassical approximation for $C$ is easier to 
obtain starting from a phase space representation, in particular 
for smooth potentials. 
We derive expressions for $C({\bf q^{+}},{\bf q^{-}})$ 
depending on the averaging procedure for essentially any given spatial 
separation $|{\bf q^{+}}-{\bf q^{-}}|$. 
In Section \ref{sec:conclusion} we relate our findings with previous 
analytical and numerical results discussing their validity range. 
We also include three appendices. In Appendix \ref{app:demDequality} 
we show the equivalence between our expression for $C$ with the one 
derived in Ref. \onlinecite{Hortikar98} in a certain limiting regime. 
Appendices \ref{app:PoleStructure} and \ref{app:IntegralC} are 
technical and devoted to the demonstration of some specific formula
appearing in the main text.

%
\section{The spectral Wigner function and its semiclassical approximation}
\label{sec:Wigner}

The Wigner function \cite{Wigner32} of an individual energy eigenstate 
$|n\rangle$ is defined by the Weyl transformation
\beq
\label{eq:defwignerfunc}               
W_n({\bf x})=
   \int \! \frac{d\mbox{\boldmath $\xi$}_{{\bf q}}}{(2\pi\hbar)^d}\; 
	   \psi_n({\bf q}^+)\psi_n^*({\bf q}^-)\,
	   \exp\left(-\frac{i}{\hbar}\mbox{\boldmath $\xi$}_{{\bf q}}
		      \cdot{\bf p}\right) \;,
\eeq
where the coordinate $\mbox{\boldmath $\xi$}_{{\bf q}}={\bf q}^+-
{\bf q}^-$, ${\bf x}=({\bf q}, {\bf p})$ is a shorthand notation for 
the phase space point with ${\bf q}=({\bf q}^++{\bf q}^-)/2$ and $d$ 
stands for the number of degrees of freedom of the system.
>From Eq.\ (\ref{eq:defwignerfunc}) it follows immediately that, upon 
averaging over the coordinate space $\langle \cdots \rangle_{\bf q}$,
the inverse Weyl transformation of $W_n({\bf x})$ gives the two-point 
autocorrelation function 
\bea
\label{eq:Ccorrfunc}
C({\bf q^{+}},{\bf q^{-}})& \equiv &
 \langle \psi_n({\bf q}^+)\psi_n^*({\bf q}^-) \rangle_{\bf q} 
\nonumber\\
&= & \int\! d{\bf p}\; 
  \left\langle W_n\!\left(\frac{{\bf q}^{+}+{\bf q}^{-}}{2},
			     {\bf p}\right) \right\rangle_{\bf q}\,
  \exp\left[\frac{i}{\hbar}\,{\bf p}
       \cdot({\bf q}^{+}-{\bf q}^{-})\right] \;.
\eea
Note that our definition of $C$ is not normalized by a factor
$\langle |\psi_n({\bf r})|^2 \rangle_{\bf q}^{-1}$ as standard.

The essence of Berry's pioneer work \cite{Berry77} was to assume
that the averaged Wigner function of a generic chaotic quantum state
$|n\rangle$ of an autonomous system is distributed as a 
Dirac $\delta$-function over the surface of energy $E_n$, {\it i.e.} 
$\langle W_n({\bf x})\rangle_{\bf q} \propto \delta[H({\bf x})-E_n]$, 
where $H({\bf x})$ is the system Hamiltonian.
For a Hamiltonian of the form
\beq
\label{eq:Hparticular}
H({\bf x})={\bf p}^2/2m + V({\bf q}) \;,
\eeq
this so-called microcanonical probability density in a $d$-dimensional 
configuration space leads to the well known formula \cite{Berry77}
\beq
\label{eq:corrbessel}
C({\bf q^{+}},{\bf q^{-}}) \propto
\frac{J_{d/2-1}\Big[p({\bf q})|{\bf q}^{+}-{\bf q}^{-}|/\hbar\Big]}
   {\Big[p({\bf q})|{\bf q}^{+}-{\bf q}^{-}|/\hbar\Big]^{d/2-1}}  \;,
\eeq
where $p({\bf q})=\sqrt{2m[E-V({\bf q})]}$, ${\bf q}=({\bf q}^{+}+
{\bf q}^{-})/2$, $J_{\nu}(x)$ is the Bessel function of order $\nu$.
(The formula encountered in Ref. \onlinecite{Berry77} differs from 
Eq.(\ref{eq:corrbessel}) by a constant due to normalization.)
Since the above relation does not depend on any system specific features,
scaling only with the local momentum $p({\bf q})$,
it directly reveals an universal behavior of chaotic wave functions.
More recently, Prigodin and collaborators \cite{super} microscopically 
obtained the same result for disordered systems using the
nonlinear $\sigma$-model, in a regime resembling quantum chaotic
systems.
Impressive numerical tests of Eq.\ (\ref{eq:corrbessel}) were 
presented, for instance in Refs.\onlinecite{Li94,Backer01}, 
reporting on the study of wave functions corresponding to high 
lying energy levels of the two-dimensional ($d=2$) conformal 
billiard.

In spite of the success of Eq.\ (\ref{eq:corrbessel}), one proviso 
ought to be made.
Both analytical and numerical results only corroborate Berry's
conjecture if one regards the average $\langle \ldots \rangle_{\bf q}$ 
in a broader sense.
The nonlinear sigma model approach averages over an ensemble of 
different impurities configurations.
In addition to the average over ${\bf q}=({\bf q}^{+}+{\bf q}^{-})/2$ 
covering regions encompassing several de Broglie wavelengths,
the local averages $\langle\ldots\rangle_{\bf q}$ 
in Ref.~\onlinecite{Li94} had to be taken over all directions of 
$({\bf q}^{+}-{\bf q}^{-})$ for a fixed value of 
$|{\bf q}^{+}-{\bf q}^{-}|$ to verify Eq.\ (\ref{eq:corrbessel}).
 
Some time ago Berry \cite{Berry89,notasaula} formulated a more 
rigorous approach to this subject, which was recently further 
developed by Ozorio de Almeida \cite{Almeida98}.
It has been shown that Berry's original conjecture of microcanonical 
probability concentration \cite{Berry77} is semiclassically verified 
if the average runs over Wigner functions of states belonging to an 
energy window containing several levels \cite{Berry89,notasaula,Almeida98}.
This construction is best casted in terms of the spectral Wigner 
function, namely
\beq
\label{eq:defspecWigfun}
W({\bf x},E;\varepsilon)\equiv
    (2\pi\hbar)^d \sum_n \delta_{\varepsilon} (E-E_n)\, W_n({\bf x}) \;,
\eeq
where the $\{E_n\}$ are the system eigenenergies. The energy smoothing 
function $\delta_\varepsilon$ is for convenience chosen as
\beq
\label{energysmooth}
  \delta_{\varepsilon} (E-E_n)=
	  \frac{\varepsilon/\pi}{(E-E_n)^2+\varepsilon^2}\;,
\eeq
in correspondence to an energy window of width $\varepsilon$
centered at $E$.
Likewise, we introduce the smoothed eigenstate
autocorrelation function 
\beq
\label{defCF}
  C_{\varepsilon}({\bf q^{+}},{\bf q^{-}},E)= \Delta
     \sum_n \delta_{\varepsilon} (E-E_n) \,
	    \psi_n({\bf q}^{+})\psi_n^*({\bf q}^{-}) \;,
\eeq
where $\Delta$ is the local mean level density, defined as $\Delta \equiv
\langle \sum_{n}\delta_\varepsilon (E- E_n)\rangle_\varepsilon^{-1}$,
with the average taken over the energy levels contained by the energy 
window $\varepsilon$ centered at $E$.
(When using the semiclassical approximation, for the sake of consistency, 
$\Delta$ is obtained from the Weyl energy level density, {\it i.e} 
$\Delta\equiv 1/\rho_{\mbox{\scriptsize W}}$.)
As follows from Eq.\ (\ref{eq:defwignerfunc}), the inverse of $W_n$ 
directly gives $\psi_n({\bf q}^{+})\,\psi_n^*({\bf q}^{-})$ rendering
\beq
\label{eq:startapp}
 C_{\varepsilon}({\bf q^{+}},{\bf q^{-}},E)= \Delta
  \int\!\frac{d{\bf p}}{(2\pi\hbar)^d}\;
  W\!\left(\frac{{\bf q}^{+}+{\bf q}^{-}}{2},{\bf p},E;\varepsilon\right)\,
  \exp\!\left[\frac{i}{\hbar}\;{\bf p}\cdot({\bf q}^{+}-{\bf q}^{-})\right]
  \;.
\eeq
The advantage of using $W$ is that it provides ways for amenable
semiclassical approximations in different energy smoothing regimes,
allowing for the analysis of $C_{\varepsilon}$ at any given spatial 
scale separation.
Eq.\ (\ref{eq:startapp}) is the starting point of all results derived 
in this paper.
The remaining of this section is devoted to the presentation of
the limiting approximations to $W$ based on the semiclassical
picture of chord and centers, postponing to the 
forthcoming section the corresponding analysis of $C_{\varepsilon}
({\bf q^{+}},{\bf q^{-}},E)$.

%
\subsection{The semiclassical spectral Wigner function}

The spectral Wigner function is related to the Weyl propagator
$U_t({\bf x})$, {\it i.e.} the Weyl transform of the propagator 
$\langle {\bf q}^+ | \exp(-i t \hat{H}/\hbar) | {\bf q}^- \rangle$,
through \cite{Berry89,Almeida98}
\beq
\label{eq:Wexact}
W({\bf x}, E;\varepsilon) = 
\frac{1}{\pi\hbar} \mbox{Re} \int_0^\infty \! dt
\;e^{-\varepsilon t/\hbar}\;U_t({\bf x})\;
\exp\left(\frac{i}{\hbar} E t \right) 
\;.
\eeq
The semiclassical approximation of $W$ is directly obtained by 
inserting in the above equation the semiclassical expression 
for the Weyl propagator \cite{Almeida98}, namely
\beq
\label{eq:weylpropagsc}
U_t^{\rm sc}({\bf x})=
\sum_j \frac{2^d}{\Big|\det [ \one + {\cal{M}}_j({\bf x},t)] 
\Big|^{1/2}}
\exp\!\left[i\;\frac{S_{j}({\bf x},t)}{\hbar}\;+\;
i\;\beta_j\right] 
\;.
\eeq
Here the sum is taken over all classical trajectories with the same 
traversal time $t$ whose phase space endpoints ${\bf x}_j^{\pm}$ 
are joined by a chord $\mbox{\boldmath $\xi$}_j({\bf x})=
{\bf x}_j^{+} - {\bf x}_j^{-}$ centered at ${\bf x}$.
Figure \ref{fig:chord_structure} illustrates the phase space
structure beneath the semiclassical Weyl propagator. 
The Maslov phase associated with the $j$-th classical trajectory 
is given by $\beta_j$.
In Eq.\ (\ref{eq:weylpropagsc}), $\one$ is the identity and 
${\cal M}_j$ is the symplectic matrix (or stability matrix).
The latter corresponds to the map 
$\delta {\bf x}_j^{+}={\cal M}_j \delta {\bf x}_j^{-}$, resulting 
from the linearization of the dynamics in the neighborhood of 
${\bf x}_j^{\pm}$.
The symbol $S_j({\bf x}, t)$ stands for the action, also called 
{\it center action} \cite{Almeida98}, given by
\beq
\label{eq:centeractiont}
S_j({\bf x}, t) = \oint_j d{\bf q}\cdot {\bf p} - \int \!dt \,
\mbox{H}[ {\bf x}_j(t)]  \;,
\eeq
where $\mbox{H}$ is the Weyl Hamiltonian, {\it i.e.} the Weyl symbol
of the Hamiltonian operator.
It is worth recalling that the Weyl Hamiltonian, $\mbox{H}(\bf x)$, 
only coincides with the classical one, $H(\bf x)$, when the latter 
is of the form given by Eq.\ (\ref{eq:Hparticular}).
The first integral at the r.h.s. of Eq.\ (\ref{eq:centeractiont}) 
is the symplectic area enclosed by the circuit taken along the 
$j$-th trajectory connecting ${\bf x}_j^{-}$ to ${\bf x}_j^{+}$ and 
closed by the chord 
$-\mbox{\boldmath $\xi$}_j({\bf x})$ (see Fig. \ref{fig:chord_structure}).
For autonomous systems the second integral is simply the product 
of the energy $E_j$ corresponding to the $j$-th trajectory, 
with its traversal time $t_j$.
The variation of $S_j$ with respect to the independent variables
\cite{Almeida98}, leads to
\beq
\label{eq:1devS}
\mbox{\boldmath $\xi$}_{j{\bf q}}= -\partial S_j/\partial{\bf p}\;,
\qquad
\mbox{\boldmath $\xi$}_{j{\bf p}}= \partial S_j/\partial{\bf q}\;,
\qquad \mbox{and} \qquad
-E_j=\partial S_j/\partial t \;.
\eeq
%
At sufficiently short times for each phase space point ${\bf x}$ 
there is only one small chord contributing to the sum in  
Eq.\ (\ref{eq:weylpropagsc}).
The short trajectory connecting the chord end points has
a Maslov phase $\beta_0=0$.
In distinction, as $t$ is increased, due to bifurcations there is 
a proliferation of different chords to be summed in 
Eq.\ (\ref{eq:weylpropagsc}).

The starting point for the semiclassical analysis of $W$
is encountered by replacing Eq.\ (\ref{eq:weylpropagsc}) 
into Eq.\ (\ref{eq:Wexact})
\beq
\label{eq:Wintermediario}
W({\bf x}, E;\varepsilon) = \frac{2^{d+1}}{2\pi\hbar} 
\mbox{Re} \sum_j \int_0^\infty \! dt
\frac{e^{-\varepsilon t/\hbar}}
{\Big|\det [ \one + {\cal{M}}_j({\bf x},t)] 
\Big|^{1/2}}
\exp\left\{\frac{i}{\hbar}[S_j({\bf x}, t) + Et + 
i \beta_j]\right\} \;.
\eeq
This formula exemplify the general structure of the semiclassical Weyl 
representation of any quantum object as being given in terms of 
its classical chord structure in phase space.
In general, this fact is revealed by the use of the 
stationary phase approximation to obtain the dominant contributions 
for any observable.
This premise will guide our analysis of $C_{\varepsilon}$ in the next
section.    
In the following, we show how the stationary phase method works 
in the case of the spectral Wigner function given by Eq.\ 
(\ref{eq:Wintermediario}).

The points of stationary phase are the solutions of
\beq
   \frac{d}{dt}\left[ S_j ({\bf x}, t) + Et\right] = 0
\;.   
\eeq
This equation fixes the traveling time along the $j$-th trajectory $t_j(E)$
at the energy $E$, for which 
\beq
\label{eq:centeractionE}
	S_j ({\bf x}, t_j) + E t_j(E) = S_j({\bf x}, E)
\eeq
where the action $S_j({\bf x}, E)$ is the symplectic area corresponding 
to the first integral at the r.h.s. of Eq.\ (\ref{eq:centeractiont}), 
with the momentum $|{\bf p}|$ fixed by the energy $E$.
In other words, the stationary phase condition selects those trajectory 
segments which belong to a single energy shell ${\cal C}$. 
Thus, for such trajectories, all the chords 
centered at ${\bf x}$ have their tips on $\cal C$.  
If all the stationary points $t_j(E)$ are isolated, which is generally
the case when the chords centered at ${\bf x}$ are sufficiently
separated, we can evaluate $W$ by stationary phase \cite{Almeida98}, 
that yields
\beq
\label{eq:scWfarC}
W({\bf x},E;\varepsilon)=\frac{2^{d+1}}{\sqrt{2\pi\hbar}}
\sum_j \;e^{-\;\varepsilon t_j/\hbar}
A_j({\bf x},E)\;
\cos\!\left[\frac{S_j({\bf x},E)}{\hbar}+\gamma_j\right] \;,
\eeq 
where the amplitude is explicitly written as
\beq
\label{eq:amplitude}
A_j({\bf x},E)= \left|\frac{d t_j}{dE} \,\det 
\!\Big\{\one + {\cal{M}}_j[{\bf x},t_j(E)]\Big\}^{-1}\right|^{1/2}    
\;,
\eeq
and we collected the Maslov phases of the classical contributions 
in $\gamma_j$.
If the energy shell $\cal C$ is closed and convex and ${\bf x}$ 
lies inside it there will be always contributing chords to Eq.\ 
(\ref{eq:scWfarC}).
To keep the presentation simple we shall only consider convex energy 
shells in this paper.

%
\subsection{The role of the energy average}

The smoothing $\varepsilon$ parameter plays an essential role in 
regulating the convergence of the semiclassical approximation for 
the spectral Wigner function $W$:
as $\varepsilon$ becomes smaller longer classical paths start 
contributing relevantly to the sum in Eq.\ (\ref{eq:scWfarC}).
It is customary to define two characteristic semiclassical scales for
$\varepsilon$ \cite{Berry89,notasaula}.
The first one is the outer scale $\varepsilon_{\rm large}\equiv \hbar/
\tau_{\rm min}$, where $\tau_{\rm min}$ is the period of the shortest 
periodic orbit, characterizing the typical time to flow around the energy 
shell. 
The second one is the inner scale $\varepsilon_{\rm small}\equiv \hbar/
\tau_{\rm H}$, where $\tau_{\rm H}$ is the Heisenberg time 
defined by the mean level spacing $\Delta$ in the considered energy 
window as $\tau_{\rm H} = \hbar/\Delta$.

Particularly strong contributions to $W$ arise when ${\bf x}$
is taken in the neighborhood of caustics. 
At such singular points the standard stationary phase approximation 
is bound to fail.  
When the evaluating point ${\bf x}$ approaches a caustic of the 
integrand in Eq.\ (\ref{eq:Wintermediario}), generically two or 
more stationary phase points $t_j(E)$ coalesce and so do the 
corresponding chords $\mbox{\boldmath $\xi$}_j({\bf x})$.
Therefore, we shall also often speak of coalescing chords at 
caustics.
The most important kind of caustics influencing $C_\varepsilon$ will 
be those at the energy shell itself and the ones near periodic orbits 
on $\cal C$.
As we will show, the first ones are associated with short times,
whereas the others with the long time dynamics.
Correspondingly we distinguish two energy averaging regimes in $W$ for 
points $\bf x$ near the energy shell:
(a) $\varepsilon \gg \varepsilon_{\rm large}$ when the signatures
of all the long trajectories are suppressed and (b) the 
opposite situation when $\varepsilon < \varepsilon_{\rm large}$ 
and more trajectories do contribute. 
Here, to keep the approximation under control,
$\varepsilon \gg \varepsilon_{\rm small}$ is required.

Let us first consider the case when $\varepsilon \gg 
\varepsilon_{\rm large}$.
Here, only one short classical trajectory, with traversal time
$t_0$ fixed by the stationary phase condition, contributes to 
$W$ \cite{Almeida98}.
The shorter chord in Fig.\ \ref{fig:chord_structure} serves to
illustratate this situation.
As ${\bf x}\rightarrow {\cal C}$, $t_0$ approaches the lower 
integration limit in Eq.\ (\ref{eq:Wintermediario}) spoiling 
the stationary phase approximation.
This difficulty can be circumvented in the following way.
Since the action $S({\bf x},t)$ is always an odd function of 
$t$ and by changing the cutoff function, $\exp(-\varepsilon t/\hbar)$, 
for an even one with respect to $t$, we write Eq.\
(\ref{eq:Wintermediario}) as
\beq
\label{eq:Wshorttraj}
W({\bf x}, E;\varepsilon) \simeq \frac{2^{d}}{2\pi\hbar} 
\int_{-\infty}^{\infty} \! dt
\frac{e^{-\varepsilon |t|/\hbar}}
{\Big|\det [ \one + {\cal{M}}_0({\bf x},t)] 
\Big|^{1/2}}
\exp\left\{\frac{i}{\hbar}[S_0({\bf x}, t) + Et]\right\} \;.
\eeq
The resulting integrand displays two stationary phase points located at 
$\pm t_0$ that coalesce at $t = 0$ as ${\bf x}$ approaches $\cal C$.
The structure of coalescing stationary points at the origin can be 
obtained by expanding the center action up to third order 
in $t$ \cite{Almeida98}
\beq
\label{eq:Supordert3}
S({\bf x},t)\approx -t \; \mbox{H}({\bf x}) -
\frac{1}{24}\,t^3\,\dot{\bf x}\,{\cal H}\,\dot{\bf x} 
\;, 
\eeq
where $\dot{\bf x}$ is the phase space velocity and ${\cal H}$ the Hessian 
matrix of the Weyl Hamiltonian, both taken at the phase space point ${\bf x}$.
The integral in Eq.\ (\ref{eq:Wshorttraj}) can now be evaluated by the 
uniform approximation method \cite{Berry76} by invoking a suitable 
change of the integration variable.
Such transformation, $z\equiv z(t)$, is the one that reduces 
the integrand phase in Eq.\ (\ref{eq:Wshorttraj}) to the canonical 
form $z^3/3 - \gamma^2 z$ 
(see, for instance,  Ref.\onlinecite{Bleistein86}). 
Mapping the stationary points $\pm t_0$ into the new ones in 
$z$, yields to $\gamma=-[3S({\bf x},E)/2]^{1/3}$. Thus, we obtain
\beq
\label{eq:scWcloseC}
W({\bf x}, E;\varepsilon) = \frac{2^{d+1}}{\sqrt{2\hbar}} 
e^{-\varepsilon |t_0|/\hbar} A_0({\bf x},E)
\left[\frac{3 S_0({\bf x}, E)}{2\hbar} \right]^{1/6} \!
\mbox{Ai} \!\left[ -\left(\frac{3 S_0({\bf x}, E)}{2\hbar} 
\right)^{2/3}\right] \;,
\eeq
where $\mbox{Ai}(y)$ is the Airy function \cite{Abramowitz64}.
This result corrects for a small mistake in Eq.\ (7.20) of 
Ref. \onlinecite{Almeida98}. It also contains the relation
previously discussed for ${\bf x}$ taken deep inside the energy 
shell.
As the evaluation point ${\bf x}$ recedes from $\cal C$,
the Airy function argument (in modulus) grows very fast. 
Hence, we are entitled to employ the Airy function approximation 
for large negative arguments and retrieve Eq.\ (\ref{eq:scWfarC}),
provided that for the short trajectory $\gamma_0=0$.

As ${\bf x}$ further approaches the energy shell $\cal C$, 
the spectral Wigner function becomes very simple (provided
$\varepsilon \gg \varepsilon_{\rm large}$).
In such situation there is an apparent indeterminacy in the amplitude 
of Eq.\ (\ref{eq:scWcloseC}) since the symplectic area $S({\bf x},E)$ 
vanishes and the amplitude $A_0({\bf x},E)$ diverges.
In this case it is an accurate approximation to represent 
the short trajectory by the short chord $\mbox{\boldmath $\xi$}_0
\approx t_0 \dot{\bf x}$, and hence the stability matrix ${\cal M}_0$ 
becomes the identity. Thus
\beq
S_0({\bf x},E)\simeq \frac{1}{12}
t_0^3\,\dot{\bf x}\,{\cal H}_0\,\dot{\bf x}=
\frac{4}{3}\sqrt{2}\,\frac{[E- \mbox{H}({\bf x})]^{3/2}}{
(\dot{\bf x}\,{\cal H}_0\,\dot{\bf x})^{1/2}}
\;,
\eeq
and
\beq
\left|\frac{dt_0}{dE}\right|^{1/2}=
\left(\frac{t_0}{2}\big|\dot{\bf x}{\cal H}\dot{\bf x}\big|
\right)^{-1/2}
\;.
\eeq
Hence, as it was already shown \cite{Berry89b,Almeida98}
\beq
W({\bf x},E;\varepsilon)\;\;_{\stackrel{\displaystyle \longrightarrow}
{\scriptscriptstyle {\bf x}\rightarrow {\cal C}}}\;\;
\frac{2}{|\hbar^2\,\dot{\bf x}\,{\cal H}_0\,\dot{\bf x}|^{1/3}}
\mbox{Ai} \!\left\{\frac{2[\mbox{H}({\bf x})-E]}
{(\hbar^2\,\dot{\bf x}\,{\cal H}_0\,\dot{\bf x})^{1/3}}\right\}
\;.
\eeq
In this situation, ${\bf x}\rightarrow {\cal C}$ and $\varepsilon \gg 
\varepsilon_{\rm large}$, for the strict semiclassical regime we easily 
recover the microcanonical probability distribution 
\beq
\label{eq:wignerespcdelta}
W({\bf x},E;\varepsilon)\approx\delta[\mbox{H}({\bf x})-E] \;,
\eeq
by recalling that $\lim_{\alpha\rightarrow 0}\alpha^{-1} 
\mbox{Ai}({\bf y}/\alpha) = \delta({\bf y})$.
This result does not come as a surprise, it is just telling 
us that we washed out most quantum interference effects  
reaching the classical limit while taking $\varepsilon \gg 
\varepsilon_{\rm large}$.
It is only by narrowing $\varepsilon$ that one can explore the rich 
structure of the spectral Wigner function and unveil non trivial
quantum features.
This discussion shall be resumed in a deeper level in the following 
section, but we can already anticipate that Eq.\ 
(\ref{eq:wignerespcdelta}) is remarkably robust.

Let us discuss now the case where $\varepsilon < \varepsilon_{\rm large}$ 
and ${\bf x}$ is taken close to the energy shell $\cal C$.
Here, one also has to account for pairs of coalescing chords in 
Eq.\ (\ref{eq:Wintermediario}), schematically shown in 
Fig.\ \ref{fig:long_traj}.
These are the short chords corresponding to long trajectories
orbiting between its tips and winding very closely to a periodic orbit.  
When ${\bf x} \rightarrow {\cal C}$ their traversal times
and actions, $S({\bf x},E)$, become degenerate with those of the 
corresponding periodic orbit.
Thus, Eq.\ (\ref{eq:scWcloseC}) has to be corrected by adding
the so called scar contributions to the spectral Wigner function,
first developed by Berry \cite{Berry89} and latter refined by
Ozorio de Almeida \cite{Almeida98}. 
The latter formula describes the Wigner scars as a peak of extra intensity
along the periodic orbits on the energy shell decorated by a fringe pattern.
As it was shown in Ref. \onlinecite{Toscano01} such Wigner scars
extend deep inside the energy shell where the spectral Wigner function 
is semiclassically given by Eq.\ (\ref{eq:scWfarC}) for any arbitrary 
value of $\varepsilon$. 
Indeed, when the action of a periodic orbit is Bohr quantized 
the contributions of trajectory segments for chords whose tips lie 
on the periodic orbit, can be added in phase to Eq.\ (\ref{eq:scWfarC}).
Hence the Wigner scars have an enhanced pattern of concentric rings 
of oscillatory amplitude on a two-dimensional surface defined by 
the centers of all the chords with endpoint on the periodic orbit
\cite{Toscano01}. 
Only the edge of this surface corresponds to the domain 
of the Berry's scars formula.
This shows that the semiclassical spectral Wigner function 
is in general not restricted to the energy shell, and thus to
short chord contributions. 
In other words, we say that the old microcanonical conjecture 
of Voros and Berry \cite{Voros76,Berry77}, has important semiclassical 
corrections.
The distinction between contributions of large and short
chords will appear again in the study of the spectral 
autocorrelation function $C_{\varepsilon}$ in the next section.

%
\section{CORRELATIONS OF ERGODIC WAVE FUNCTIONS FOR DIFFERENT
	 ENERGY AVERAGING REGIMES.}
\label{sec:nonuniversal}

In this section we derive a general semiclassical formula
for $C_\varepsilon ({\bf q^{+}},{\bf q^{-}},E)$ expressed 
in terms of the system classical chord structure.
Our analysis is build on the semiclassical approximations 
for the spectral Wigner function $W$ presented in the foregoing 
section. 

For any given pair of points in position space, ${\bf q}^{+}$ and 
${\bf q}^-$, the integral in Eq.\ (\ref{eq:startapp}) is performed 
over the entire $d$-dimensional momentum space.
In this study we only consider ${\bf q}^+$ and ${\bf q}^-$ within 
classically allowed regions.
Hence, the integration momentum space intercepts the energy 
shell and is naturally divided into a domain located 
in the interior of $\cal C$  and another at its exterior.
This is illustrated in Fig.\ \ref{fig:chord_C}.
For convex energy surfaces, the ones considered here, the spectral 
Wigner function, whose argument is ${\bf q}=({\bf q}^+ 
+{\bf q}^-)/2$, exponentially vanishes for values of ${\bf p}$ in 
the phase space region exterior to $\cal C$.
Hence, in general, the main contribution to the integral in 
Eq.\ (\ref{eq:startapp}) arises from momenta in the interior 
of $\cal C$ and at its immediate neighborhood.
Thus, we are allowed to bound the effective momentum integration space 
in Eq.\ (\ref{eq:startapp}) to the classically allowed momenta.
We name the so defined integration space the ``momentum space 
associated to ${\bf q}=({\bf q}^+ +{\bf q}^-)/2$".
 
In line with Section \ref{sec:Wigner}, first we find the general 
chord structure that $C_{\varepsilon}$ inherits from $W$.
This is done by inserting into Eq.\ (\ref{eq:startapp})
the semiclassical expression for the 
spectral Wigner function given by Eq.\ (\ref{eq:scWfarC}), which is
valid for any arbitrary $\varepsilon$.
The resulting integral can be casted as
\beq
\label{eq:CFI1I2} 
C_{\varepsilon}({\bf q^{+}},{\bf q^{-}},E)=
\Delta
\sum_j 
\frac{e^{-\varepsilon t_j/\hbar}}{(2\pi\hbar)^{d+1/2}}
\left[
    I_{j}^+({\bf q}^{+},{\bf q}^{-},E)\,e^{i\gamma_j}+
    I_{j}^-({\bf q}^{+},{\bf q}^{-},E)\,e^{-i\gamma_j}
						     \right]  \;,
\eeq
with
\beq
\label{eq:defI1I2}
I_{j}^\pm =2^{d}
\int \! d{\bf p} \, 
A_j\left({\bf x},E\right)
\exp\left\{\frac{i}{\hbar}
\left[
S_j\left({\bf x},E\right)
\pm {\bf p}\cdot({\bf q}^{+}-{\bf q}^{-})
\right]
\right\}
\;.
\eeq
Now we evaluate $I_j^\pm$ by stationary phase. The stationary phase 
points ${\bf p}_j\equiv{\bf p}_j ({\bf q}^{+},{\bf q}^{-},E)$ are 
solutions of 
\beq
\label{eq:stphase1}
\frac{\partial}{\partial {\bf p}}
\left[
S_j\left(\frac{{\bf q}^{+}+{\bf q}^{-}}{2},{\bf p},E\right)
\right]=\mp({\bf q}^{+}-{\bf q}^{-})
\eeq
for $I_{j}^\pm$. 
We recall that the variation of $S_j({\bf x},E)$ with respect
to the independent variables ${\bf q}$ and ${\bf p}$ are 
given in Eq.\ (\ref{eq:1devS}).
Hence, the first member of Eq.\ (\ref{eq:stphase1}) is exactly 
$-\mbox{\boldmath $\xi$}_{j{\bf q}}=-({\bf q}_j^{+}-{\bf q}_j^{-})$.
In other words, the stationary phase condition selects those classical 
$j$-trajectories in the energy shell whose projected chords in the 
configuration space match the vectors $\pm({\bf q}^{+}-{\bf q}^{-})$.
This geometrical structure is sketched in Fig. \ref{fig:chord_C}, where
the projected chords are indicated by dashed vectors.
In the same figure, the classical $j$-trajectories are those flowing 
between the intersections of the momentum spaces corresponding to 
${\bf q}={\bf q}^-$ and ${\bf q}={\bf q}^+$ with the energy surface 
${\cal C}$.
The panel (a) of Fig. \ref{fig:chord_C} corresponds to time reversal 
symmetric flows, whereas (b) represents the cases when this symmetry 
is absent.
The locations of the many possible stationary phase points
${\bf p}_j$ in the momentum space associated to ${\bf q}=({\bf q}^{+}
+{\bf q}^-)/2$ are indicated by dots and astericks.
While for ${\bf q}^{-}\rightarrow {\bf q}^{+}$ the chords centered 
at ``astericks'' coalesce to a zero length chord at the energy shell 
$\cal C$, the ones centered at ``dots''  are typically large.
This allow us to name the chords centered at ``astericks'' as ``short'' 
chords and that ones centered at ``dots'' as ``long'' chords.

Provided that the stationary points ${\bf p}_j$ are sufficiently 
far apart from each other we can safely evaluate $I_j^+$ and $I_j^-$ 
by the stationary phase method. 
The latter requires a symmetrized Legendre transformation in the phases
of Eq.\ (\ref{eq:defI1I2}), which is conveniently expressed by 
the standard textbook action $\mbox{S}$ \cite{Almeida98}, with variables 
in the configuration space, namely
\beq
\label{eq:Sq1q2legendreT}
\mbox{S}_j({\bf q}^{\pm},{\bf q}^{\mp},E)=
S_j({\bf x}_j,E)
\pm {\bf p}_j\cdot 
({\bf q}^{\pm}-{\bf q}^{\mp}) \;,
\eeq
where ${\bf x}_j=({\bf q}, {\bf p}_j)$ with ${\bf q} = ({\bf q}^{+}+
{\bf q}^{-})/2$ and ${\bf p}_j \equiv {\bf p}_j({\bf q}^{+},
{\bf q}^{-},E)$.
We obtain for the non oscillatory factor of both $I_j^{+}$ 
and $I_j^{-}$ 
\beq
\label{EQ:AMPLITUDEI}
2^d A_j({\bf x}_j,E)
   \left|\det\left(
   \frac{\partial^2 S_j({\bf x},E)}{\partial {\bf p}^2}\right)
   \right|^{-1/2}\Bigg|_{{\bf x}={\bf x}_j} =
|D_j|^{1/2}
\;,
\eeq
where  
\beq
\label{eq:Dequality}
D({\bf q}^{+},{\bf q}^{-},E)\equiv (-1)^d
  \det\left(
  \matrix{ \frac{\partial^2\mbox{S}}{\partial{\bf q}^-\partial{\bf q}^+} 
	&  \frac{\partial^2\mbox{S}}{\partial{\bf q}^-\partial{E}}  \cr
	   \frac{\partial^2\mbox{S}}{\partial{\bf q}^+\partial{E}}
	&  \frac{\partial^2\mbox{S}}{\partial E^2}  \cr} \right)
\;.   
\eeq 
The details of the derivation leading to Eq.\ (\ref{EQ:AMPLITUDEI}) can
be found in Appendix \ref{app:demDequality}.

For time-reversal invariant systems, to every $j$-trajectory on 
$\cal C$ going from ${\bf q}^-$ to ${\bf q}^+$ (solution of the 
integral $I_j^{+}$) there is a corresponding time reversed pair 
going from ${\bf q}^+$ to ${\bf q}^-$ (solution of the integral 
$I_j^{-}$).
Evidently both terms contribute with the same stationary phase 
and amplitude to the sum in Eq.\ (\ref{eq:CFI1I2}). 
Hence
\beq
\label{eq:hortikarresult}
C_{\varepsilon}({\bf q^{+}},{\bf q^{-}},E)\approx
\frac{2\,\Delta}{(2\pi\hbar)^{(d+1)/2}}
\sum_j
e^{-\;\varepsilon t_j/\hbar}
|D_j({\bf q}^{+},{\bf q}^{-},E)|^{1/2}
\cos\left[\frac{\mbox{S}_j({\bf q}^{+},{\bf q}^{-},E)}{\hbar}
+ \gamma_j + \nu_j \frac{\pi}{4} \right]
\;,
\eeq
where $\nu_j\equiv\mbox{sgn}[\partial^2 S_j({\bf x},E)/\partial 
{\bf p}^2]$. 
This is essentially the main finding of Ref. \onlinecite{Hortikar98}.

Let us now discuss the conditions under which Eq.\ 
(\ref{eq:hortikarresult}) fails.
When ${\bf q}^-$ approaches ${\bf q}^+$, the points ${\bf x}_j$, 
represented by astericks in Fig. \ref{fig:chord_C}, move closer to the 
energy shell.
As we learned in Section \ref{sec:Wigner} this is a case where
the semiclassical approximation for the spectral Wigner function, 
from which Eq.\ (\ref{eq:hortikarresult}) was obtained, fails. 
Remarkably, even in this limit where ${\bf q}^- \rightarrow {\bf q}^+$
there are ``long'' chords whose center points ${\bf x}_j$, 
indicated by dots in Fig. \ref{fig:chord_C}, are typically far 
from $\cal C$.
Such contributions to $C_\varepsilon$ are still well 
described by Eq.\ (\ref{eq:hortikarresult}), but are obviously 
unrelated to the ``short'' chords.  
The approximation scheme for $W$ developed in the previous section 
to deal with the classical contributions due to the ``short'' chords 
centered at points ${\bf x}_j$ close to  ${\cal C}$, can be also 
used to obtain $C_\varepsilon$.
In the remaining of this section we pursue this path for
two very different energy smoothing regimes and obtain a semiclassical 
approximation of $C_{\varepsilon}$ valid for any spatial separation scale
within the classical allowed region. 

At this point an important remark is in order.
The chord structure for a fixed distance $|{\bf q}^+ - {\bf q}^-|$
sketched in Fig. \ref{fig:chord_C} is robust upon changes of 
${\bf q}=({\bf q}^+ + {\bf q}^-)/2$ unless one of the points 
${\bf q}^{\pm}$ reaches the boundary of the classical
allowed region.
When this condition is met, the ``short'' chords coalesce with the
``long'' ones, corresponding in Eq.\ (\ref{eq:CFI1I2}) to the 
case of coalescing stationary points.
A semiclassical investigation for a similar situation
was already reported \cite{Berry89b}, but it did not address 
wave function correlations.
This is a most interesting physical situation because of its relation 
to tunneling rates and possible implications to the already mentioned 
Coulomb blockade systems 
\cite{Alhassid97,Vallejos99,Narimanov99,Kaplan00,Narimanov01}.
Unfortunately we were not able to develop a semiclassical approximation
to this problem yet. We avoid it in this paper by restricting our 
analysis to points ${\bf q}^{\pm}$ inside the classical allowed region 
and distant from its boundary by a couple of de 
Broglie wavelengths.

\subsection{Large $\varepsilon\gg\varepsilon_{\rm large}$ smoothing
            regime}
\label{subsec:e>emax}

For $\varepsilon\gg\varepsilon_{\rm large}$ the cutoff parameter 
$\varepsilon$ suppresses all but the shortest trajectory contribution 
to Eq.\ (\ref{eq:hortikarresult}). 
The latter connects both tips of a ``short'' chord $\mbox{\boldmath 
$\xi$}_{0}$ (see Fig. \ref{fig:chord_C}).
As discussed before, when the center ${\bf x}_0$ of 
$\mbox{\boldmath $\xi$}_{0}$ approaches the energy shell ${\cal C}$, 
we have to employ the corresponding uniform approximation to $W$, 
given by Eq.\ (\ref{eq:scWcloseC}).
 
Instead of directly Fourier transforming the semiclassical 
spectral Wigner function, it is advantageous to step back, use $W$ as 
given by Eq.\ (\ref{eq:Wshorttraj}) and invert the integration order. 
That is
\beq
\label{eq:startappCe>>emax}
C_{\varepsilon}({\bf q^{+}},{\bf q^{-}},E) \simeq 
  \frac{\Delta}{2\pi\hbar}\int_{-\infty}^{\infty} \! dt\;
      e^{-\varepsilon |t|/\hbar} \, F_0({\bf q}^+,{\bf q}^-,t)
      \exp\!\left(\frac{iEt}{\hbar}\right) \;,
\eeq
where 
\beq
\label{eq:Kq1q2}
F_0({\bf q}^+,{\bf q}^-,t) \equiv \frac{2^{d}}{(2\pi\hbar)^d}
\int \! d{\bf p} \frac{1}
{{\Big|\det [ \one + {\cal{M}}_0({\bf x},t)] 
\Big|^{1/2}}}
\exp\!\left\{\frac{i}{\hbar}\Big[S_0({\bf x}, t) + {\bf p} \cdot
 ({\bf q}^+ - {\bf q}^-)\Big]\right\} .
\eeq
The above integral is evaluated by the stationary phase method.
The stationary phase point ${\bf p}_0\equiv{\bf p}_0({\bf q}^+,
{\bf q}^-,t)$ is the solution of
\beq
\label{eq:stphaseR0}
-\mbox{\boldmath $\xi$}_{0{\bf q}}\equiv
\frac{\partial}{\partial {\bf p}}
\left[
S_0\left(\frac{{\bf q}^{+}+{\bf q}^{-}}{2},{\bf p},t\right)
\right]=-({\bf q}^{+}-{\bf q}^{-}) \;.
\eeq
In analogy to the case that lead to Eq.\ (\ref{eq:Sq1q2legendreT}),
the phase factor is best written in terms of the standard text book 
action $\mbox{R}$, with variables in the configuration space, namely
\beq
\label{eq:Rq1q2legendreT}
\mbox{R}_0({\bf q}^{+},{\bf q}^{-},t)= S_0({\bf x}_0,t)
       +{\bf p}_0\cdot({\bf q}^{+}-{\bf q}^{-})
\;.
\eeq
Substituting the obtained stationary phase approximation of $F_0$
into Eq.\ (\ref{eq:startappCe>>emax}) we write $C_\varepsilon$ as
\beq
\label{eq:Cq1q2e>>emax}
C_{\varepsilon}({\bf q^{+}},{\bf q^{-}},E) \simeq
\frac{\Delta}{(2\pi\hbar)^{d/2+1}}\int_{-\infty}^{\infty} \! dt\;
g(t)\,
\exp\!\left[\frac{i}{\hbar}\;
\Phi(t,\theta)+i\nu_0(t)\frac{\pi}{4}\right] \;,
\eeq
where $\nu_0(t)\equiv\mbox{sgn}[\partial^2 S_0({\bf x},t)/\partial 
{\bf p}^2]$.
The function $g(t)$ gives the amplitude
\beq
\label{eq:funcg(t)}
g(t)=2^d\;e^{-\varepsilon |t|/\hbar}\;
\left|\det \Big[ \one + {\cal{M}}_0({\bf x},t)\Big] 
\det\!\left(
\frac{\partial^2 S_0({\bf x},t)}{\partial {\bf p}^2}
\right)
\right|^{-1/2}\Biggr|_{{\bf x}={\bf x}_0}
\;,
\eeq
with ${\bf x}_0=({\bf q},{\bf p}_0)$, ${\bf q} = ({\bf q}^{+}+
{\bf q}^{-})/2$ and ${\bf p}_0 \equiv {\bf p}_0({\bf q}^{+},
{\bf q}^{-},t)$. 
The phase $\Phi$ stands for
\beq
\label{eq:phasePhi}
\Phi(t,\theta)=\mbox{R}_0({\bf q}^{+},{\bf q}^{-},t) +Et \;,
\eeq
where we introduced $\theta\propto |{\bf q}^+-{\bf q}^-|$ as 
a control parameter of the integral, {\it i.e.}
${\bf x}_0\rightarrow {\cal C}$ when $\theta\rightarrow 0$.

The integral in Eq.\ (\ref{eq:Cq1q2e>>emax}) is dominated by the 
stationary phase points located at $\pm t_0(\theta,E)$ 
corresponding to the traversal time of the shortest trajectory 
going from ${\bf q}^-$ to ${\bf q}^+$ and the one running backwards 
in time.
As $\theta\rightarrow 0$, $\pm t_0(\theta,E)$ coalesce 
at the origin. 
This situation again is very similar to the one encountered in the
previous section when dealing with the spectral Wigner function
for ${\bf x}$ near the energy shell [{\it i.e.} that one that 
leads to Eq.\ (\ref{eq:scWcloseC})].
The difference is in the functional form of the phase $\Phi$
and on the behavior of $g(t)$ near the origin.
For a Hamiltonian of the form Eq.\ (\ref{eq:Hparticular}) we show 
in Appendix \ref{app:PoleStructure} that both the phase $\Phi$ 
and the amplitude $g(t)$ have a singularity at the origin.
Notwithstanding, the integral in Eq.\ (\ref{eq:Cq1q2e>>emax}) is 
finite and can be evaluated using the uniform approximation method 
\cite{Bleistein86}. 
The result is
\beq
\label{EQ:CBRIDGE}
C_{\varepsilon}({\bf q^{+}},{\bf q^{-}},E)=
\frac{2\pi\,\Delta}{(2\pi\hbar)^{\frac{d}{2}+1}}\;
\Big[|D_0({\bf q}^+,{\bf q}^-,E)|\;
\mbox{S}_0({\bf q}^+,{\bf q}^-,E)\Big]^{1/2}
J_{\frac{d}{2}-1}
\left[\frac{\mbox{S}_0({\bf q}^+,{\bf q}^-,E)}{\hbar}\right]
\;,
\eeq
where we left out the smoothing factor $e^{-\varepsilon |t_0|/\hbar}$,
since for practical purposes the condition $t_0 \ll \hbar/\varepsilon$ 
is always met.
Details about the evaluation of the integral in Eq.\ 
(\ref{eq:Cq1q2e>>emax}) leading to (\ref{EQ:CBRIDGE}) are found in 
Appendix \ref{app:PoleStructure}.

The above semiclassical approximation for $C_{\varepsilon}({\bf q^{+}},
{\bf q^{-}},E)$ is valid for any separation $|{\bf q}^+-{\bf q}^-|$,
provided the arguments belog to the classical allowed region.
Let us examine the small and large separation limits.
As ${\bf q}^- \rightarrow {\bf q}^+$ the shortest trajectory on 
$\cal C$ is well approximated by the ``short'' chord 
$\mbox{\boldmath $\xi$}_{0}$, and the action $\mbox{S}_0$ turns 
into
\beq
\label{EQ:APPROXFORS0}
\mbox{S}_0({\bf q^{+}},{\bf q^{-}},E)\approx
	       p_0({\bf q})\,|{\bf q}^+-{\bf q}^-|
\;,
\eeq
where $p_0({\bf q})=\sqrt{2m[E-V({\bf q})]}$. Hence, the determinant 
$D_0$ simplifies to
\beq
\label{EQ:APPROXFORD0}
|D_0({\bf q}^+,{\bf q}^-,E)|^{1/2}\approx
m\; \frac{p_0({\bf q})^{(d-3)/2}}{
|{\bf q}^+-{\bf q}^-|^{(d-1)/2}}
\;.
\eeq
(The derivation of Eqs.\ (\ref{EQ:APPROXFORS0}) and (\ref{EQ:APPROXFORD0})
is found in Appendix \ref{app:IntegralC}.)
Collecting the results, we write
\beq
\label{eq:correbessel}
C({\bf q^{+}},{\bf q^{-}})=
\frac{(2\pi)^{d/2}\, m\; p({\bf q})^{d-2}}{(2\pi\hbar)^{d}\,
\rho_{\mbox{\scriptsize W}}(E)} \;
\frac{J_{d/2-1}\Big[p({\bf q})|{\bf q}^{+}-{\bf q}^{-}|/\hbar\Big]}
   {\Big[p({\bf q})|{\bf q}^{+}-{\bf q}^{-}|/\hbar\Big]^{d/2-1}}  
\eeq
and recover, by an appropriate normalization, Berry's original result 
Eq.\ (\ref{eq:corrbessel}).
In the corresponding opposite limit, when $|{\bf q}^+-{\bf q}^-|$ is
large, we use the asymptotic expansion of the Bessel function for 
large arguments, namely, $J_{\nu}(x)\approx \sqrt{2/(\pi x)}\cos(x-
\nu\pi/2-\pi/4)$ and retrieve the semiclassical approximation given by 
Eq.\ (\ref{eq:hortikarresult}) for the shortest trajectory.

\subsection{Small $\varepsilon<\varepsilon_{\rm large}$ smoothing regime}
\label{subsec:e<emax}

As the smoothing parameter $\varepsilon$ is shrinked, longer 
trajectories have to be taken into account.
As a consequence, when $\varepsilon < \varepsilon_{\rm large}$, 
the approximation scheme becomes subtler than the simplified 
one discussed in Section \ref{subsec:e>emax}.
Before exploring this regime, it is useful to remind ourselves that 
the semiclassical contributions to the spectral Wigner function 
come from orbits connecting the tips of either ``short" or ``long" 
chords, as depicted in Fig. \ref{fig:chord_C}.
We thus classify the classical trajectories into three categories. 
(a) Trajectories connecting tips of ``long" chords. We already
discussed this case at the beginning of Section \ref{sec:nonuniversal}. 
Here, since the trajectories are isolated, we just use Eq.\ 
(\ref{eq:hortikarresult}). Difficulties, if any, arise from the 
large number of trajectories entering the sum, as regulated by
the $\varepsilon$.
(b) Short trajectories connecting ``short chords". Those were
analyzed in the foregoing subsection. 
(c) Long trajectories connecting ``short chords".
The analysis of $C_{\varepsilon}({\bf q^{+}},{\bf q^{-}},E)$ becomes 
much clearer in the two cases discussed in the following, 
namely, ${\bf x}$ close and far from the energy surface ${\cal C}$.

When the long trajectories have tips on ``short'' chords, and 
${\bf q}^-$ approaches ${\bf q}^+$, the centers ${\bf x}_j$
of these chords approach $\cal C$. 
Here in addition to the trajectories of type (a) and (b), 
we have to account for the long trajectories associated with ``short'' 
chords of the type (c) but whose centers are close to $\cal C$.
Hence, as already discussed, for evaluating points close
to the energy shell we must use a suitable $W$ for Eq.\ 
(\ref{eq:startapp}).
This is just Berry's scar expansion formula in terms of periodic 
orbits \cite{Berry89,notasaula,Almeida98}.
Hence, for small distances $|{\bf q}^+ - {\bf q}^-|$, we 
write
\beq
\label{eq:Ce<emax-shortgeral}
C_{\varepsilon}({\bf q^{+}},{\bf q^{-}},E)=
\sum_{\stackrel{j \in {\scriptsize \mbox{``long"}}}
	         {\scriptsize \mbox{chords}}} 
C_{\varepsilon}^j({\bf q^{+}},{\bf q^{-}},E) +
C_{\varepsilon}^0({\bf q^{+}},{\bf q^{-}},E) +
C_{\varepsilon}^{\scriptsize \mbox{scar}}({\bf q^{+}},{\bf q^{-}},E)
\;,
\eeq
where $C_{\varepsilon}^j$ stands for the $j$-th type (a) orbit 
corresponding to a term in the sum in Eq. \ (\ref{eq:hortikarresult}), 
$C_{\varepsilon}^0$ for the type (b) trajectory term given by 
Eq.\ (\ref{EQ:CBRIDGE}), and $C_{\varepsilon}^{\rm scar}$ 
is the result of Eq.\ (\ref{eq:startapp}) for $W$ taken as the scar 
expansion formula. 
We do not intend to present here any detailed analysis of 
this integration, but it is easy to realize that the final result 
will be an expansion in terms of all the periodic orbits 
that pass through the points ${\bf q}^{\pm}$. 
Such conjecture is further supported by noting that when 
${\bf q}^-={\bf q}^+= {\bf q}$ the integration Eq.\ 
(\ref{eq:startapp}) reduces to the projection `down' 
${\bf p}$ of the Berry's scar expansion.
The latter case corresponds to Bogomolny formula \cite{Bogomolny88}
for scars in the probability density in configuration space, which 
involves all the periodic orbits that pass through the point ${\bf q}$ 
(see Ref. \onlinecite{notasaula}). 

In the other limit, when $|{\bf q}^+ - {\bf q}^-|$ is large, the 
centers ${\bf x}_j$ of all the chords sketched in Fig. \ref{fig:chord_C} 
are far from $\cal C$, and we thus write 
\beq
\label{eq:Ce<emax-longgeral}
C_{\varepsilon}({\bf q^{+}},{\bf q^{-}},E)=
\sum_{j \neq 0}
C_{\varepsilon}^j({\bf q^{+}},{\bf q^{-}},E) +
C_{\varepsilon}^0({\bf q^{+}},{\bf q^{-}},E)
\;,
\eeq
where we separate the contribution $C_{\varepsilon}^0$ of the
shortest trajectory given by the uniform expression 
Eq. \ (\ref{EQ:CBRIDGE}).

It is interesting to note that both formulas Eq.\ 
(\ref{eq:Ce<emax-shortgeral}) and (\ref{eq:Ce<emax-longgeral}) 
imply that the spectral autocorrelation function has contributions 
arising from $W$ taken at points ${\bf x}_j$  well inside the 
energy shell.
In the case of Eq.\ (\ref{eq:Ce<emax-shortgeral}) these are 
the trajectories with tips on ``long'' chords.
In general, these contributions can not be neglected,
as already discussed at the end of Section \ref{sec:Wigner}.
Particularly, the ``off-shell'' scars of the spectral
Wigner function \cite{Toscano01}
provide the only periodic orbit contributions 
for $C_{\varepsilon}$ in the case of large separation $|{\bf q}^+ 
- {\bf q}^-|$.

\subsection{Spatial averaging with $\varepsilon<\varepsilon_{\rm large}$}
\label{subsec:spatial_averaging}

We now investigate the effect of spatial averaging on the
spectral autocorrelation function.
We are interested to know under which circumstances this averaging 
washes out the system specific features, allowing one to define a 
local system independent (or universal) regime for 
$C_{\varepsilon}$.
Furthermore, the additional spatial averaging brings our results 
into close relation to numerical experiments, such as the ones in 
Refs. \onlinecite{Li94,Backer01}. 

Defining the local spatial average as
\beq
\label{eq:mediaq}
\langle C_{\varepsilon}({\bf q}^+,{\bf q}^-,E)\rangle_{\bf q}
\equiv
\frac{1}{A(\Omega)}\int_{\Omega}d{\bf q}\; 
C_{\varepsilon}({\bf q}^+,{\bf q}^-,E)
\;,
\eeq
where $\Omega$ is a configuration space region of surface
$A(\Omega)$ (in $d$-dimension), covering many ``local'' de 
Broglie wavelengths $\lambda_{\bf q}$ across.
We define $\lambda_{\bf q} =2\pi\hbar/p({\bf q})$ in terms of
the local momentum $p({\bf q})$, consistent with the semiclassical 
approximation.
The average is restricted to a region $\Omega$ of classically 
small variations of the smooth potential $V({\bf q})$.

The ubiquitous robustness of Berry's expression for the autocorrelation 
of chaotic wave functions can be attributed to the following: 
At small separations $|{\bf q}^+ - {\bf q}^-|$ the local space average 
kills the terms $C_{\varepsilon}^j$ associated with the ``long'' chords.
This suppression is due to the fact that such terms oscillate 
in a scale smaller (or at most comparable) to the de Broglie 
wavelength.
Indeed, for small separation the ``long'' chords are almost parallel to
the momentum space and thus its centers have 
${\bf p}_j({\bf q}^+,{\bf q}^-,E)\approx 0$ (see Fig. \ref{fig:chord_C}). 
>From Eq.\ (\ref{eq:Sq1q2legendreT}), $\mbox{S}_j({\bf q}^+,{\bf q}^-,E)
\approx S_j({\bf x}_j,E)$ and thus the local spatial wavelength in 
Eq.\ (\ref{eq:hortikarresult}) is approximately
\beq
\label{eq:wavelenghtLjq}
\lambda_{j{\bf q}}
\stackrel{\scriptscriptstyle j\neq 0}{\approx}
2\pi\hbar
\left|
\frac{\partial S_j({\bf x}_j,E)}{\partial {\bf q}}
\right|^{-1}
=\frac{2\pi\hbar}{|\mbox{\boldmath $\xi$}_{j{\bf p}}|}
\;,
\eeq
This is just about the de Broglie wavelength since 
$|\mbox{\boldmath $\xi$}_{j{\bf p}}|\sim 2\,p({\bf q})$.  
Moreover, the requirement of small spatial averaging regions
assures that $p({\bf q})$ is approximately constant for ${\bf q}$ 
inside $\Omega$. 
Consequently, $C_{\varepsilon}^0$ for small $|{\bf q}^+ - {\bf q}^-|$ 
coincides with Berry's result and remains unaffected by the local 
spatial average \cite{comentario}.
Furthermore, if none of the wave functions inside the energy 
window from which $C_{\varepsilon}$ is built shows a strong visual 
scar due to periodic orbits, which is often the case, there is no
reason to expect a sizeable correction due to $\langle C_{
\varepsilon}^{\scriptsize \mbox{scar}}\rangle_{\bf q}$. 
Thus, after the averaging the leading contribution for 
$C_{\varepsilon}$ would be the local system independent expression
Eq. \ (\ref{eq:corrbessel}). 

Extrapolating our semiclassical analysis of $\langle C_{\varepsilon}
({\bf q}^+,{\bf q}^-,E)\rangle_{\bf q}$ to energy smoothings of the 
order of one energy spacing ({\it i.e.}, $\varepsilon\sim\varepsilon_
{\rm small}$), we find that our results are still consistent with the 
numerical investigations \cite{Li94,Backer01} of the autocorrelation
function $C({\bf q}^+,{\bf q}^-)$ [in this case Eq.\ (\ref{eq:Ccorrfunc})]
on individual eigenfunctions for small values of $|{\bf q}^+ - {\bf q}^-|$.
For instance, in Ref. \onlinecite{Backer01} we observe that the 
corrections to Eq.\ (\ref{eq:corrbessel}) given by 
$\langle C_{\varepsilon}^{\scriptsize \mbox{scar}}\rangle_{\bf q}$
are small unless the eigenfunction has a strong visual scar
due to a simple periodic orbit.
Likewise, as Eq.\ (\ref{eq:corrbessel}) is symmetric with respect
to the orientations of $({\bf q}^+-{\bf q}^-)$ for a fixed value
of $|{\bf q}^+ - {\bf q}^-|$, the observed angular dependence 
of $C({\bf q}^+,{\bf q}^-)$ \cite{Li94,Backer01} arrives precisely
from 
$\langle C_{\varepsilon}^{\scriptsize \mbox{scar}}\rangle_{\bf q}$.

Before concluding we like to add that the local spatial average is 
the mechanism responsible for the elimination of the contributions 
of trajectories associated with ``long'' chords in the Bogomolny 
scar formula for the spatial probability density \cite{Bogomolny88}.
In our formalism, the latter is recovered after
making the local average Eq.\ (\ref{eq:mediaq}) over the autocorrelation 
function $C_\varepsilon$, Eq.\ (\ref{eq:Ce<emax-shortgeral}),
and then taking ${\bf q}^+ \rightarrow {\bf q}^-$. 
In other words, the Bogomolny scar formula captures (in configuration
space) only the scar contributions of periodic orbits near the energy 
shell of the spectral Wigner function, since the ``off-shell'' scar 
contributions \cite{Toscano01} are washed out by the local spatial 
average.


\section{Conclusions}
\label{sec:conclusion}

We investigated the spatial two-point autocorrelation of energy 
eigenfunctions $\psi_n({\bf q})$ corresponding to classically 
chaotic systems in the semiclassical regime.
We use the Weyl-Wigner formalism to obtain the spectral average 
$C_{\varepsilon}({\bf q^{+}},{\bf q^{-}},E)$ 
of $\psi_n({\bf q}^{+})\psi_n^*({\bf q}^{-})$, defined as
the average over eigenstates within an energy window $\varepsilon$
centered at $E$.
In the considered framework $C_{\varepsilon}$ 
is just the Fourier transform in momentum space of the spectral
Wigner function $W({\bf x},E;\varepsilon)$.

The advantage of this formalism comes from the observation that $W$
is almost like tailormade for semiclassical approximations.
At each phase space point, ${\bf x}\equiv({\bf q},{\bf p})$,
the semiclassical behavior of $W$ is associated with all the classical 
trajectories on the energy shell $E$ whose end points are joined by a 
chord centered at ${\bf x}$. 
These classical contributions are exponentially suppressed when the 
trajectory traversal time is $t \stackrel{\scriptscriptstyle >}
{\scriptscriptstyle\sim}\hbar/\varepsilon$.

In distinction to most studies so far, this paper addresses 
smooth Hamiltonian systems of the form given by Eq.\ ({\ref{eq:Hparticular}). 
Our results can evidently be straightforwardly  employed  to 
calculate $C_{\varepsilon}$ in billiard systems, where the
phase space structure is much simpler than the one considered here.
For smooth systems, we show that it is still possible to distinguish 
in $C_{\varepsilon}$ between a local system independent regime and
another one that carries the system classical chord structure 
information.
This is obviously also valid for billiards. 
The interplay between the spectral average window, which controls the 
upper time scale of the classical contributions, and the spatial 
separation scale dictates which aspects prevails.
As a result we obtain semiclassical expressions that bridge the 
existing formulae for the autocorrelation function $C_\varepsilon$.

To the best of our knowledge, the studies of $C_\varepsilon$
found in the literature are based on Green's function methods, 
and employ a given arbitrary separation between ``zero-length" and 
``long" trajectories. 
In billiards, due to their simplicity, the semiclassical 
``zero-length" Green's function is an excellent approximation 
to calculate $C_\varepsilon$, even for $\varepsilon$ comparable 
with $\varepsilon_{\scriptsize \mbox{\rm large}}$, provided that
the separation is not too large.
In smooth systems corrections accounting for the energy surface 
curvature become rapidly necessary as the spatial separation is
increased.
Such corrections, albeit in principle feasible to obtain with the 
Green's function method, are easier to estimate with the here employed 
framework. 
This is an important advantage of the formalism we employ.
Our study goes beyond that issue showing that the Wigner-Weyl formalism 
is a quite general framework for semiclassical approximations, clearly 
revealing the inextricable relation between the classical chord 
structure and the choatic wave function correlations.

\acknowledgements
We thank Alfredo M. Ozorio de Almeida for his several insightful 
remarks and suggestions which enormously contributed to this study. 
One of the authors (FT) acknowledges the financial support
by FAPERJ and CLAF/CNPq (Brazil). 
This work was supported in part by CNPq and PRONEX (Brazil).

\appendix

\section{Demonstration of Eq.\ (\ref{EQ:AMPLITUDEI})}
\label{app:demDequality}

In this Appendix we show that both amplitudes in Eq.\ 
(\ref{EQ:AMPLITUDEI}) are identical, thus proving that 
Ref. \onlinecite{Hortikar98} addresses one of the limiting 
cases of $C_\varepsilon$ studied in this paper.
For the sake of clarity, it is convenient to express both sides of 
Eq.\ (\ref{EQ:AMPLITUDEI}) in terms of the action defined as
$\mbox{R}({\bf q}^+,{\bf q}^-,t)=\mbox{S}({\bf q}^+,{\bf q}^-,E)-Et$.
The variation of $\mbox{R}$ with respect to its independent variables,
namely, ${\bf q}^+, {\bf q}^-$, and $t$ gives \cite{Gutzwiller90}
\beq
\label{eq:1devR}
{\bf p}^{\pm}=\pm\partial \mbox{R}/\partial{\bf q}^{\pm} 
\qquad \mbox{and} \qquad
-E=\partial \mbox{R}/\partial t
\;.
\eeq
It is also convenient to introduce a short notation for the 
second derivatives of R
\bea
\label{eq:2devR}
\mbox{R}_{++}\equiv\frac{\partial^2 \mbox{R}}{\partial{\bf q}^{+2}}=
\frac{\partial{\bf p}^+}{\partial {\bf q}^+}  \qquad  
\mbox{R}_{+-}\equiv\frac{\partial^2 \mbox{R}}{\partial {\bf q}^+ 
\partial{\bf q}^-}=-
\frac{\partial{\bf p}^-}{\partial {\bf q}^+} &
\nonumber \\
\mbox{R}_{-+}\equiv\frac{\partial^2\mbox{R}}{\partial{\bf q}^-
\partial{\bf q}^+}=
\frac{\partial{\bf p}^+}{\partial {\bf q}^-} \qquad
\mbox{R}_{--}\equiv\frac{\partial^2\mbox{R}}{\partial {\bf q}^{-2}}=-
\frac{\partial{\bf p}^-}{\partial {\bf q}^-} &
\;,
\eea
which form a set of four $d$-dimensional matrices.

With the elements at hand one readily writes $D({\bf q}^+,
{\bf q}^-,E)$, as defined in Eq.\ (\ref{eq:Dequality}) by
a $(d + 1)\times (d+1)$-dimensional matrix determinant
(see, for instance, Ref. \onlinecite{Gutzwiller90})
\beq
\label{eq:DL+1E}
D=
\left(\frac{\partial^2\mbox{R}}{\partial t^2}\right)^{-1}
\det (-\mbox{R}_{+-})
= -
\left(\frac{\partial t}{\partial E}\right)
\det (-\mbox{R}_{+-}) \;.
\eeq

We switch now to the l.h.s. of Eq.\ (\ref{EQ:AMPLITUDEI}).
The $2d \times 2d$-dimensional stability matrix
${\cal M}({\bf x})$ is related to the second derivatives
of the action $\mbox{R}$ as follows
\beq
\label{eq:relMwithR}
{\cal M}({\bf x})=
\left(
\matrix{  
- \mbox{R}^{-1}_{+-} \mbox{R}_{--}  &  -\mbox{R}^{-1}_{+-} \cr 
\mbox{R}_{-+}-\mbox{R}_{++}\mbox{R}^{-1}_{+-}\mbox{R}_{--} &   
- \mbox{R}_{++} \mbox{R}^{-1}_{+-} \cr 
} 
\right)\;.
\eeq
The justification for the above equation can be found in the
Appendix A of Ref. \onlinecite{Berry89}.
It is now straightforward to show that
\beq
\label{eq:det1M}
\det\big[\one + {\cal M}({\bf x})\big]=
\frac{\det(\mbox{R}_{+-}-\mbox{R}_{++}-\mbox{R}_{--}+\mbox{R}_{-+})}
{\det(-\mbox{R}_{+-})}\;.
\eeq

It remains only to express $\partial^2S/\partial {\bf p}^2$ as a 
function of R to conclude the demonstration.
The relation between the second derivatives of the center
action $S({\bf q},{\bf p},E)$ and the stability matrix ${\cal M}({\bf x})$
can be obtained by differentiating ${\bf q}^+\equiv{\bf q}^+
({\bf q}^-,{\bf p}^-)$
and
${\bf p}^+\equiv{\bf p}^+({\bf q}^-,{\bf p}^-)$ with respect 
to ${\bf q}$ and ${\bf p}$.
By writing  ${\bf q}^- = {\bf q}^-({\bf q},{\bf p})$,
${\bf p}^-= {\bf p}^-({\bf q},{\bf p})$, and recalling
that
${\bf q}^{\pm}={\bf q}\mp(\partial S/\partial {\bf p})/2$
and
${\bf p}^{\pm}={\bf p}\pm(\partial S/\partial {\bf q})/2$,
we arrive at
\beq
\label{eq:2derSandM}
[\one-{\cal J}{\cal B}({\bf x})]={\cal M}({\bf x})\;
[\one+{\cal J}{\cal B}({\bf x})]\;,
\eeq
where
\beq
{\cal B}({\bf x})=\frac{1}{2}
\left(
\matrix{\frac{\partial^2S}{\partial {\bf q}^2}&  
	\frac{\partial^2S}{\partial{\bf p}\partial{\bf q}}  \cr 
	\frac{\partial^2S}{\partial{\bf q}\partial{\bf p}}  & 
	\frac{\partial^2S}{\partial {\bf p}^2}   \cr } 
\right)
\qquad \mbox{and}\qquad
{\cal J}=
\left(
\matrix{0&\one \cr 
	-\one & 0 \cr } 
\right)\;.
\eeq
Using the obvious identity ${\cal J}^{-1}=-{\cal J}$ we recast 
${\cal B}({\bf x})$ as
\beq
\label{eq:ex2derSandM}
{\cal B}({\bf x})=-{\cal J}[\one-{\cal M}({\bf x})]
[\one+{\cal M}({\bf x})]^{-1}\;.
\eeq
The above relation with the aid of Eq.\ (\ref{eq:relMwithR}) renders
\beq
\label{eq:S2vsR}
\frac{\partial^2S}{\partial {\bf p}^2}=
2^{2d}(\mbox{R}_{+-}-\mbox{R}_{++}-\mbox{R}_{--}+\mbox{R}_{-+})^{-1}
\;.
\eeq
%
%

Finally, by collecting (\ref{eq:DL+1E}), (\ref{eq:det1M}), and 
(\ref{eq:S2vsR}) we arrive at
\bea
\label{eq:expression-for-D}
-\left(
\frac{\partial t}{\partial E}
\right)&&
\left\{\det\big[\one + {\cal M}({\bf x})\big]
\det \!
\left(
\frac{\partial^2S}{\partial {\bf p}^2}
\right)
\right\}^{-1}=
\nonumber \\
&&=
-\left(
\frac{\partial t}{\partial E}
\right)
\frac{\det(-\mbox{R}_{+-})}{2^{2d}}=
\frac{D}{2^{2d}}\;,
\eea
which proves Eq.\ (\ref{EQ:AMPLITUDEI}).

\section{Derivation of the uniform approximation 
Eq.\ (\ref{EQ:CBRIDGE})}
\label{app:PoleStructure}

In this appendix we evaluate the integral Eq.\ (\ref{eq:Cq1q2e>>emax})
by the uniform approximation method, to obtain Eq.\ (\ref{EQ:CBRIDGE}) 
for the spectral autocorrelation function.
As already mentioned, the integral in Eq.\ (\ref{eq:Cq1q2e>>emax}) 
is dominated by its stationary phase points $\pm t_0(\theta,E)$ 
(solutions of $\partial\mbox{R}_0/\partial t + E=0$), that
coalesce at the origin as $\theta \rightarrow 0$. 
Thus, we start analyzing the structure of the phase $\Phi(t,\theta)$ 
and the amplitude $g(t)$ near $t=0$.

For $\Phi(t,\theta)$ we replace the action $\mbox{R}_0$ by 
the one found in Eq.\ (\ref{eq:Rq1q2legendreT}). 
We then expand the center action $S_0({\bf x}, t)$ [providing 
that ${\bf p}_0 \equiv {\bf p}_0({\bf q}^{+},{\bf q}^{-},t)$ is 
fixed by the condition Eq.\ (\ref{eq:stphaseR0})], up to third 
order in $t$, as in Eq.\ (\ref{eq:Supordert3}).
If the system Hamiltonian has the form of Eq.\ 
(\ref{eq:Hparticular}), the expansion reads 
\beq
\label{eq:Supordert3-ESP}
S_0({\bf x},t)\approx -t \; H({\bf x}) -
\frac{1}{24}\,t^3\,
\left(
\frac{1}{m}\left|\frac{\partial V}{\partial {\bf q}}\right|^2 
+
\frac{1}{m^2}\,{\bf p}\;\frac{\partial^2 V}{\partial {\bf q}^2}
\;
{\bf p}
\right)
\;, 
\eeq
so for  Eq.\ (\ref{eq:stphaseR0}) we have
\beq
\label{eq:stphaseR0-ESP}
-\frac{\partial S_0}{\partial{\bf p}}
\Biggr|_{{\bf x}={\bf x}_0}
=
\frac{t}{m}\,{\cal A}({\bf q},t)\,{\bf p}_0=({\bf q}^+-{\bf q}^-)
\;,
\eeq
where ${\bf x}_0=({\bf q},{\bf p}_0)$, with ${\bf q} = 
({\bf q}^{+}+{\bf q}^{-})/2$ and 
${\cal A}$ is the $d\times d$-dimensional matrix
\beq
\label{eq:Amatrix}
{\cal A}({\bf q},t)=\one + \frac{t^2}{12m}\,
\frac{\partial^2 V}{\partial {\bf q}^2}(\bf q)
\;.
\eeq
Solving Eq.\ (\ref{eq:stphaseR0-ESP}) for ${\bf p}_0$
and using that ${\cal A}^{-1}=\one-\frac{t^2}{12m}
\partial^2 V/\partial{\bf q}^2+{\cal O}(t^4)$, we
arrive at
\beq
\label{eq:p0func-t}
{\bf p}_0({\bf q}^{+},{\bf q}^{-},t)=
\frac{m}{t}({\bf q}^+-{\bf q}^-) -\frac{t}{12}\,
\frac{\partial^2 V}{\partial {\bf q}^2}({\bf q}) +
{\cal O}(t^3)\;,
\eeq
where $m({\bf q}^+-{\bf q}^-)/t$ comes from the first 
order term in the $t$-expansion of the center action Eq.\ 
(\ref{eq:Supordert3-ESP}).
Eqs.\ (\ref{eq:Supordert3-ESP}) and (\ref{eq:p0func-t}) yield
\beq
\label{eq:phasePhi-ESP}
\Phi(t,\theta)=\frac{m}{2t}|{\bf q}^+-{\bf q}^-|^2\;
+[E-V({\bf q})]t
+\tilde{\Phi}(t,\theta)
\;.
\eeq
Here the explicit part of $\Phi$ arises from the lowest
order in Eq.\ (\ref{eq:Supordert3-ESP}) and 
$\tilde{\Phi}(t,\theta)$ is an analytical function of $t$.
The phase $\Phi(t,\theta)$ has a pole of order
one at the origin of the complex plane $t$.

The behavior of $g(t)$, defined by Eq.\ (\ref{eq:funcg(t)}), 
near $t=0$ is dominated by
\beq
\label{eq:Mod-det-dev2S0}
\left|
\det\!\left(
\frac{\partial^2 S_0({\bf x},t)}{\partial {\bf p}^2}
\right)
\right|^{-1/2}\Biggr|_{{\bf x}={\bf x}_0}\approx
\frac{m^{d/2}}{|t|^{d/2}}\;\Big|\det 
\left[{\cal A}({\bf q},t)\Big]\right|^{-1/2}\;,
\eeq
as seen from Eq.\ (\ref{eq:stphaseR0-ESP}).
For clarity purposes we introduce $\tilde{g}(t)$ an analytical 
function of $t$, defined as $g(t)\equiv\tilde{g}(t)/|t|^{d/2}$. 

The center action $S_0$ and ${\bf p}_0({\bf q}^{+},{\bf q}^{-},t)$
are odd function of $t$ \cite{Almeida98}, and consequently so is 
the phase $\Phi(t,\theta)$. 
Since $g(t)$ is an even function of $t$, we are allowed to express 
Eq.\ (\ref{eq:Cq1q2e>>emax}) by
\beq
\label{eq:Cq1q2e>>emax-MOD}
\frac{2\Delta}{(2\pi\hbar)^{d/2+1}}\,
\mbox{Re}\; I(\hbar,\theta)\equiv 
\frac{2\Delta}{(2\pi\hbar)^{d/2+1}}\,
\mbox{Re}
\int_{0}^{\infty} \! dt\;
\frac{\tilde{g}(t)}{t^{d/2}}\,
\exp\!\left[\frac{i}{\hbar}\;
\Phi(t,\theta)-id\frac{\pi}{4}\right] \;,
\eeq
where we used that $\nu_0(t)\equiv\mbox{sgn}[\partial^2 
S_0({\bf x},t)/\partial {\bf p}^2]$, which is the difference 
between the number of positive and negative eigenvalues of 
$\partial^2 S_0({\bf x},t)/\partial {\bf p}^2$, is simply
equal to $-td/|t|$ when the energy shell is convex (the case 
considered in this study).
We have now to deal with a single stationary phase point 
$t_0(\theta, E)$, that coalesces with the lowest limit of 
integration, and for which the stationary phase reads
\beq
\label{eq:stphase-Phi}
\Phi(t_0,\theta)=\mbox{R}_0({\bf q}^{+},{\bf q}^{-},t_0) +Et_0\equiv
\mbox{S}_0({\bf q}^+,{\bf q}^-,E)
\;.
\eeq

The uniform approximation \cite{Bleistein86} to Eq.\ 
(\ref{eq:Cq1q2e>>emax-MOD}) involves the suitable change of the 
integration variable, $w\equiv w(t):[0,+\infty)\rightarrow 
[0,+\infty)$, such that it is invertible and reduces the phase 
of the integrand to the canonical form   
\beq
\label{eq:phase-canonical-form}
\Phi(t,\theta)=\frac{1}{2}\left(w+\frac{z^2(\theta)}{w}\right)
\equiv \phi(w,\theta)
\;,
\eeq
adhering to $\Phi(t,\theta)$ as given by Eq.\ (\ref{eq:phasePhi-ESP}).
Likewise, we must require that the stationary points $\pm t_0$
of $\Phi$ correspond to the saddle points $\pm w_0\equiv \pm z(\theta)$
of $\phi$ 
({\it i.e.} $\partial \phi/\partial w|_{w=\pm w_0}=0$ for
$\pm w_0=w(\pm t_0)$).
This is achieved by making $z(\theta)=\mbox{S}_0({\bf q}^+,{\bf q}^-,E)$
[see Eq.\ (\ref{eq:stphase-Phi})].
Thus, after the change of variable the integral
in Eq.\ (\ref{eq:Cq1q2e>>emax-MOD})
becomes
\beq
\label{eq:Cq1q2e>>emax-MOD-NEW}
I(\hbar,\theta)=
\int_{0}^{\infty} \! dw\;
\frac{\tilde{G}(w)}{w^{d/2}}\,
\exp\!\left[\frac{i}{\hbar}\;
\phi(w,\theta)-id\frac{\pi}{4}\right] \;,
\eeq
where
\beq
\label{eq:Gtilde}
\frac{\tilde{G}(w)}{|w|^{d/2}}=
\frac{\tilde{g}(t)}{|t|^{d/2}}
\left|\frac{\partial t}{\partial w}\right|
\;,
\eeq
and $\tilde{G}$ is an even analytical function of $w$.
In general the next step of the method of uniform approximation
is to expand $\tilde{G}$ around $w_0$ \cite{Bleistein86}, but 
in our case is sufficient to keep only the first term, namely
\beq
\label{eq:Gtilde-approx}
\tilde{G}(w)\approx\tilde{G}(w_0)=
g\Big(t_0\equiv t(w_0)\Big)
\left|\frac{\partial t}{\partial w}\right|
\Biggr|_{w=w_0}
|w_0|^{d/2}
\;,
\eeq
where
\beq
\label{eq:devt-dev-w}
\left|\frac{\partial t}{\partial w}\right|
\Biggr|_{w=w_0}=
\frac{|\partial \phi/\partial w||_{w=w_0}}
{|\partial \Phi/\partial t||_{t=t_0}}=
\frac{1}{|\mbox{S}_0({\bf q}^+,{\bf q}^-,E)|^{1/2}}
\left|\frac{\partial t_0}{\partial E}\right|^{1/2}
\;.
\eeq
The last equality was obtained by applying the L' Hospital rule.
Now, recalling
Eq.\ (\ref{EQ:AMPLITUDEI}) we write
\beq
\label{eq:Gtilde-w0}
\tilde{G}(w_0)=e^{-\varepsilon |t_0|/\hbar}
|D_0({\bf q}^+,{\bf q}^-,E)|^{1/2}\;
\mbox{S}_0({\bf q}^+,{\bf q}^-,E)^{(d-1)/2}
\;,
\eeq
and using  
an integral representation of the Hankel's function
$H^{(1)}_{\nu}(z)$ \cite{Abramowitz64}, we have
\beq
\label{eq:Hankel-func}
I(\hbar,\theta)\approx
\frac{\pi\;G(w_0)}{\mbox{S}_0({\bf q}^+,{\bf q}^-,E)^{d/2-1}}\;
H^{(1)}_{d/2-1}\left[\frac{\mbox{S}_0({\bf q}^+,{\bf q}^-,E)}
{\hbar}\right]
\;.
\eeq
Finally, collecting the results of Eqs.\ (\ref{eq:Hankel-func}) and 
(\ref{eq:Gtilde-w0}) and taking the real
part in Eq.\ (\ref{eq:Cq1q2e>>emax-MOD}) we obtain the 
uniform approximation to Eq.\ (\ref{EQ:CBRIDGE}).

\section{Demonstration of Eqs.\ (\ref{EQ:APPROXFORS0}) and 
         (\ref{EQ:APPROXFORD0})}
\label{app:IntegralC}

In this appendix we derive Eqs.\ (\ref{EQ:APPROXFORS0}) and 
(\ref{EQ:APPROXFORD0}), approximate expressions for the action 
$\mbox{S}_0({\bf q}^+,{\bf q}^-,E)$ and the determinant $D_0$
respectively. 
Both relations are valid provided the control parameter
$\theta \propto |{\bf q}^+-{\bf q}^-|$ is small.
Eqs.\ (\ref{EQ:APPROXFORS0}) and (\ref{EQ:APPROXFORD0}) show that 
our result for the spectral autocorrelation function Eq.\ 
(\ref{EQ:CBRIDGE}) reduces to Berry's one.

Let us start with Eq.\ (\ref{EQ:APPROXFORS0}). 
The action $\mbox{S}_0$ is given by Eq.\ (\ref{eq:stphase-Phi})
where $t_0$ is determined by the stationary phase condition
$\partial \mbox{R}_0/\partial t =-E$.
For $\theta$ small, $\mbox{R}_0$ is 
\beq
\label{eq:R0firstorder}
\mbox{R}_0({\bf q}^+,{\bf q}^-,t)\approx
\frac{m}{2t}\;|{\bf q}^+-{\bf q}^-|^2-V({\bf q})t
\;,
\eeq
(see Eq.\ (\ref{eq:phasePhi-ESP}) and the discussion preceding it).
The stationary phase point $t_0$ is thus
\beq
\label{eq:t0-firstorder}
t_0\approx\frac{m}{p_0({\bf q})}
|{\bf q}^+-{\bf q}^-|
\;,
\eeq
where $p_0({\bf q})=\sqrt{2m[E-V({\bf q})]}$.
By using Eq.\ (\ref{eq:R0firstorder}) and Eq.\ (\ref{eq:t0-firstorder})
in Eq.\ (\ref{eq:stphase-Phi}) we arrive to the approximation 
Eq.\ (\ref{EQ:APPROXFORS0}).

For the determinant $D_0$, from Eq.\ (\ref{eq:expression-for-D})
we write
\beq
\label{eq:D0appC}
|D_0({\bf q}^+,{\bf q}^-,E)|^{1/2}=
2^d\;
\left|\frac{\partial t_0}{\partial E}\right|^{1/2}\;
\Big|\det\big[\one + {\cal M}_0({\bf x},t_0)\big]\Big|^{-1/2}\;
\left|
\det \!
\left(
\frac{\partial^2S_0({\bf x},t_0)}
{\partial {\bf p}^2}
\right)
\right|^{-1/2}
\Biggr|_{{\bf x}={\bf x}_0}
\eeq
where ${\bf x}_0=({\bf q},{\bf p}_0)$ is the center of the
shortest chord with tips on a classical
trajectory (see Section \ref{subsec:e>emax}).
We used the center action $S({\bf x},t)$ instead of 
$S({\bf x},E)$ because for any of both $\mbox{\boldmath $\xi$}_{\bf q}
({\bf x})= -\partial S/\partial{\bf p}$. 
When $\theta \rightarrow 0$, the center ${\bf x}_0$ approaches
the energy shell, the chord $\mbox{\boldmath $\xi$}_0 \approx 
t_0 \dot{\bf x}$ [$t_0$ is given by Eq.\ (\ref{eq:t0-firstorder})] 
well approximate the classical trajectory in phase space, 
and hence ${\cal M}_0$ becomes the identity map, yielding
\beq
\label{eq:devt0-devE-approx}
\left|\frac{\partial t_0}{\partial E}\right|^{1/2}=
\left|\frac{\partial^2 R_0}{\partial t^2}\right|^{-1/2}
\Biggr|_{t=t_0}
\approx
\frac{m}{p_0({\bf q})^{3/2}}\;
|{\bf q}^+-{\bf q}^-|^{1/2}
\;,
\eeq
where we used Eq.\ (\ref{eq:R0firstorder}) for $\mbox{R}_0$.
Up to the same order considered in Eq.\ (\ref{eq:R0firstorder})
the matrix $\cal A$ in Eq.\ (\ref{eq:Mod-det-dev2S0}) becomes 
the identity and
\beq
\label{eq:det-devS0-devp2-approx}
\left|
\det \!
\left(
\frac{\partial^2S_0({\bf x},t_0)}
{\partial {\bf p}^2}
\right)
\right|^{-1/2}
\Biggr|_{{\bf x}={\bf x}_0}
\approx
\frac{m^{d/2}}{|t_0|^{d/2}}=
\frac{p_0({\bf q})^{d/2}}{|{\bf q}^+-{\bf q}^-|^{d/2}}
\;.
\eeq
Collecting these results 
in Eq.\ (\ref{eq:D0appC}) 
we arrive at Eq.\ (\ref{EQ:APPROXFORD0}).

%


%
\begin{figure}
\setlength{\unitlength}{1cm}
\begin{picture}(0,6)(0,0)
\put(2.5,-3.0){\epsfxsize=8.5cm\epsfbox[0 0 595 842]{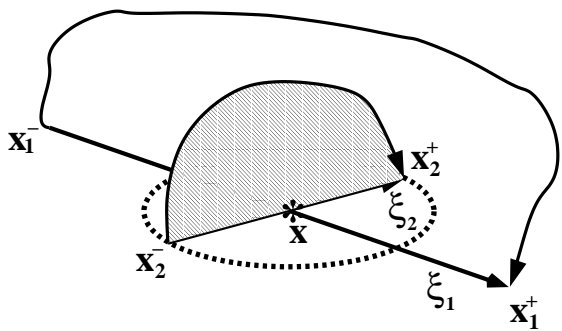}}
\end{picture} 
\caption{
Chord structure of the semiclassical Weyl propagator, Eq.\ 
(\ref{eq:weylpropagsc}), and of the spectral Wigner function,
Eq.\ (\ref{eq:Wintermediario}), in a $2d$-dimensional phase 
space.
For a central point ${\bf x}$ we show typical chords with 
their tips connected by classical trajectories on different 
energy shells but with the same traversal time $t$. 
The dashed line indicates the locus, on a given energy shell
$E$, of all the tips of chords centered at ${\bf x}$. 
The classical trajectories connecting these type of chords 
are the semiclassical contributions to $W$.
The dashed area represents the symplectic area corresponding to 
the action $S_2({\bf x}, E)$ (see text).}
\label{fig:chord_structure}
\end{figure}

%
\begin{figure}
\setlength{\unitlength}{1cm}
\begin{picture}(0,9)(0,0)
\put(3.5,-0.5){\epsfxsize=7.5cm\epsfbox[0 0 595 842]{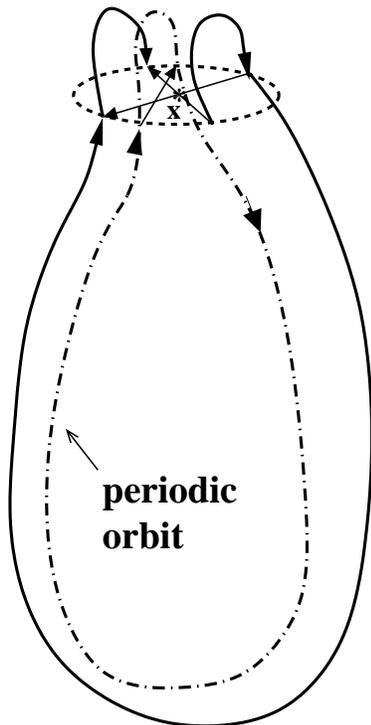}}
\end{picture} 
\caption{
Illustration of a composition of trajectory segments that 
closely follow a periodic orbit.
Berry's scar formula single out the trajectories connecting the 
tips of the two depicted chords which become indistinguishable 
from the period orbit as ${\bf x}$ approaches ${\cal C}$.} 
\label{fig:long_traj}
\end{figure}

%
\begin{figure}
\setlength{\unitlength}{1cm}
\begin{picture}(0,11)(0,0)
\put(-0.5,0.0){\epsfxsize=8.5cm\epsfbox[0 0 595 842]{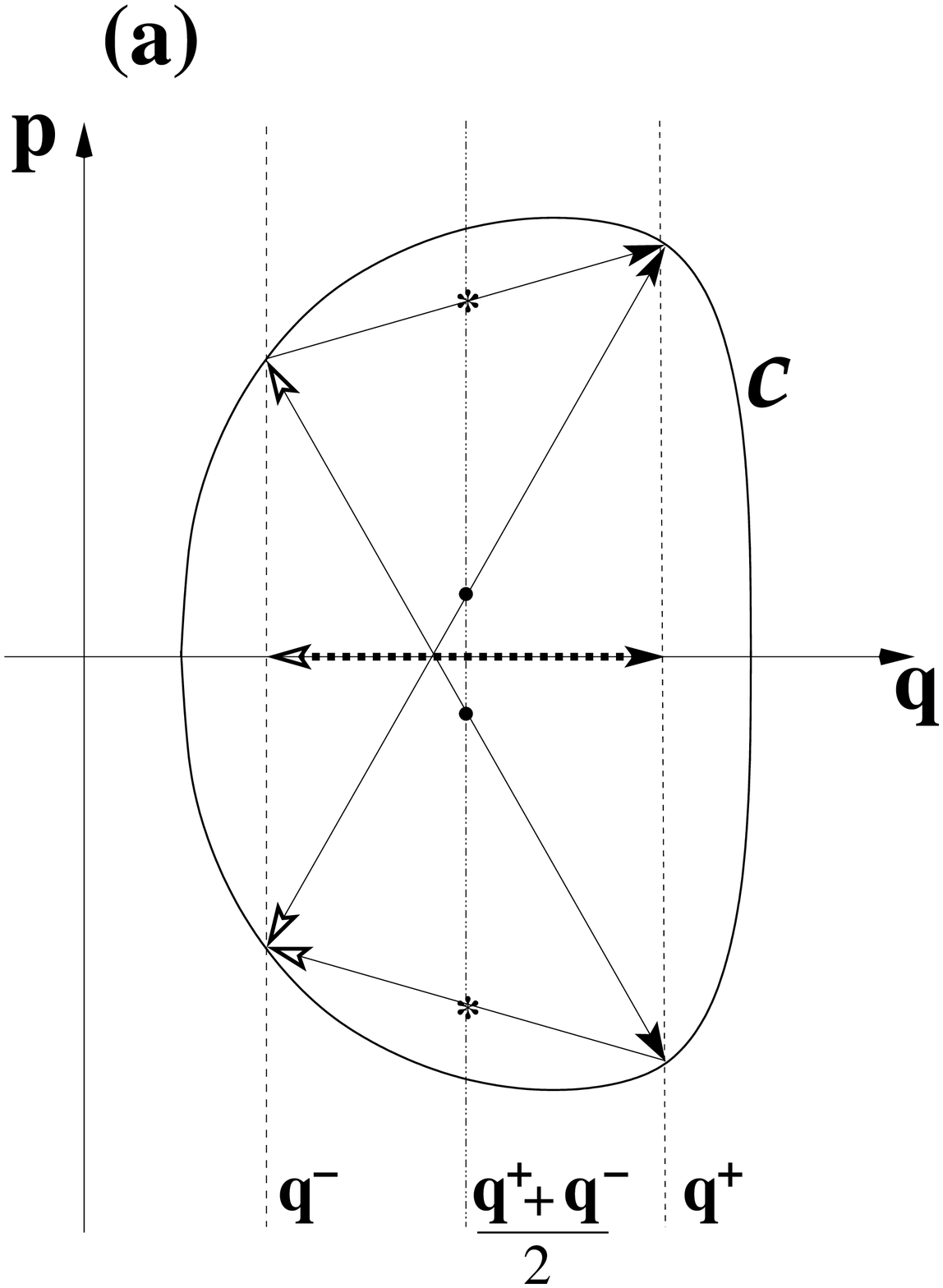}}
\put(7.5,0.0){\epsfxsize=8.5cm\epsfbox[0 0 595 842]{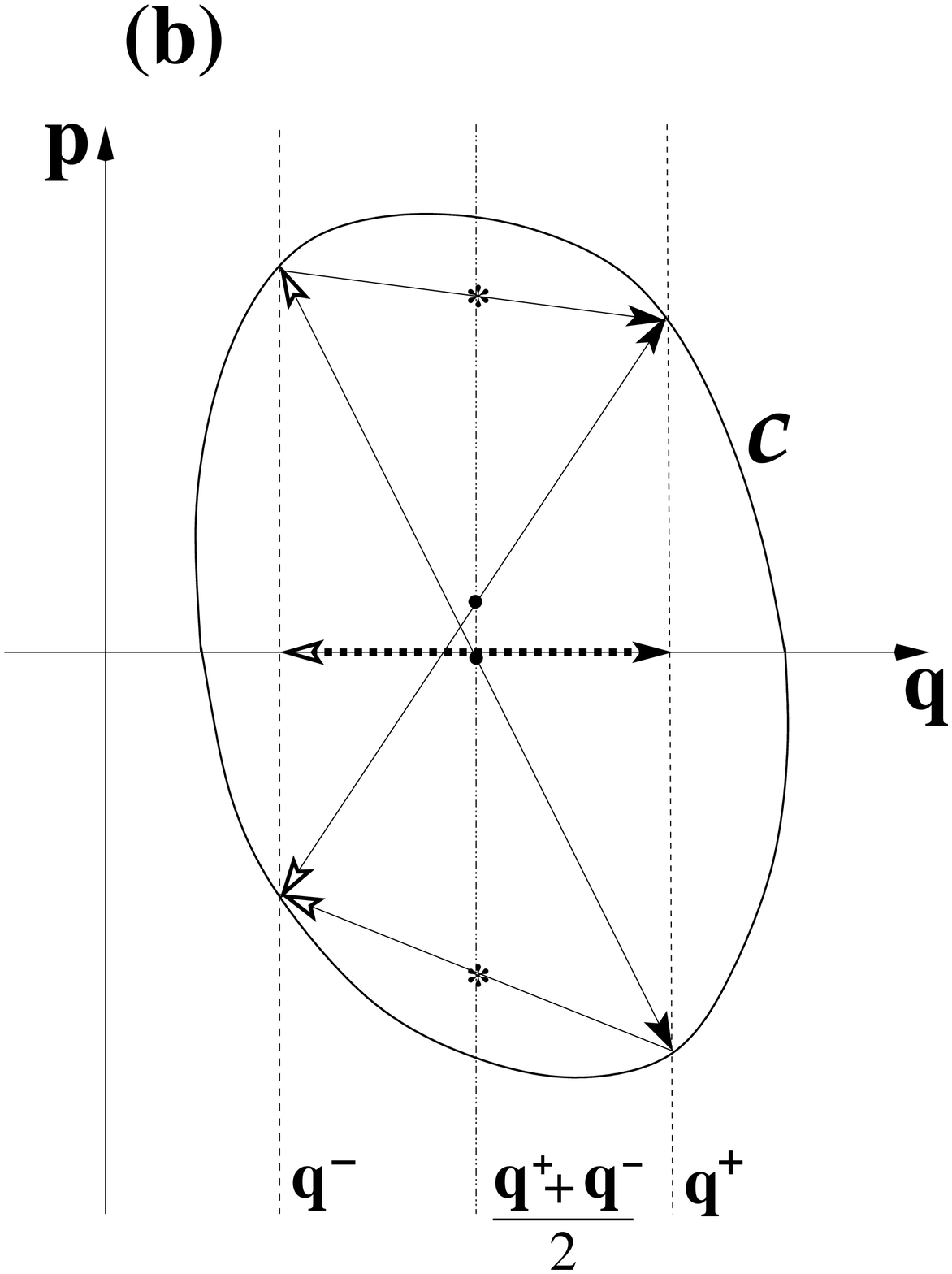}}
\end{picture} 
\caption{
Chord structure of the semiclassical approximation for the spectral 
autocorrelation function $C_{\varepsilon}$.
The horizontal dashed lines represent the projection on configuration 
space of the chords $\mbox{\boldmath $\xi$}_{\bf q}$ matching the 
vectors $\pm({\bf q}^{+}-{\bf q}^-)$.  
The points of stationary phase are located on the $d$-dimensional 
associated momentum space to ${\bf q}=({\bf q}^{+}+{\bf q}^-)/2$ 
and indicated by astericks ($\ast$) and dots ($\bullet$) (See text 
for details). 
$\cal C$ stands for the $(2d-1)$-dimensional surface of constant 
energy $E$, whereas the axis ${\bf q}$ and ${\bf p}$ represent 
$d$-dimensional surfaces.
The semiclassical contributions to $C_{\varepsilon}$ are given by
all trajectories on $\cal C$ flowing between the tips of chords
$\mbox{\boldmath $\xi$}_{\bf q}$. Panel (a) illustrates a typical 
time reversal symmetric case, while (b) a case when this symmetry
is absent.}
\label{fig:chord_C}
\end{figure}

\end{document}